\documentclass[article,onecolumn,preprint,nofootinbib,prd]{revtex4}
\pdfoutput=1
\usepackage[letterpaper,left=.75in,right=.75in,top=.75in,bottom=1.in]{geometry}
\usepackage{graphicx,amsfonts,color,comment,amsmath,hyperref,float}
\usepackage{amssymb}	
\usepackage{scrextend}

\usepackage{mathrsfs,amssymb}  
\usepackage{cancel}
\usepackage[normalem]{ulem}

\newcommand{\be}{\begin{equation}}
\newcommand{\ee}{\end{equation}}
\newcommand{\bea}{\begin{eqnarray}}
\newcommand{\eea}{\end{eqnarray}}

\usepackage{color}
\usepackage{graphicx}  
\usepackage{xcolor}

\begin{document}

\title{Investigating Dark Matter and MOND Models with Galactic Rotation Curve Data}

\date{\today} \author{Mads T. Frandsen}\email{frandsen@cp3.sdu.dk}
\affiliation{$CP^3$-Origins, University of Southern Denmark, Campusvej 55, DK-5230 Odense M, Denmark}
\date{\today} \author{Jonas Petersen}\email{petersen@cp3.sdu.dk}\affiliation{$CP^3$-Origins, University of Southern Denmark, Campusvej 55, DK-5230 Odense M, Denmark}

\begin{abstract}

We study geometries of galactic rotation curves from Dark Matter (DM) and Modified Newtonian Dynamics (MOND) models in $(g_{\rm bar},g_{\rm tot})$-space ($g2$-space) where $g_{\rm tot}$
is the total centripetal acceleration of matter in the galaxies and $g_{\rm bar}$ is that due to the baryonic (visible) matter assuming Newtonian gravity.

The $g2$-space geometries of the models and data from the SPARC database
are classified 
and compared in a rescaled $\hat{g}2$-space that reduces systematic uncertainties on galaxy distance, inclination angle and variations in mass to light ratios.

We find that MOND modified inertia models, frequently used to fit rotation curve data, are disfavoured at more than 5$\sigma$ independent of model details. The Bekenstein-Milgrom formulation of MOND modified gravity compares better with data in the analytic approximation we use. However a quantitative comparison with data is beyond the scope of the paper due to this approximation. 
NFW DM profiles only agree with a minority of galactic rotation curves. 

Improved measurements of rotation curves, in particular at radii below the maximum of the total and the baryonic accelerations of the curves are very important in discriminating models aiming to explain the missing mass problem on galactic scales.

\end{abstract}
\preprint{CP3-Origins-2018-018 DNRF90}

\maketitle

\section{introduction}
\noindent The fact that gravitational potentials on a range of astrophysical scales are deeper than predicted in Newtonian gravity is well established based on a variety of astronomical observations.
These include measurements of the rotation curves of baryonic matter in galaxies \cite{Rubin:1970zza,Rubin:1980zd,Bosma:1981zz}, the velocity dispersion of galaxies in clusters \cite{Zwicky:1933gu}, lensing of merging clusters \cite{Clowe:2006eq} and measurements of the cosmic microwave background \cite{Ade:2015xua}. 
This fact is also referred to as the "missing mass problem" and observations on all the aforemetioned scales have been argued to be in overall agreement with the presence of particle dark matter as the solution. Challenges for DM models in e.g. accounting for structure on small scales, such as the cusp-core problem \cite{Flores:1994gz}, the missing sattelites problem \cite{Moore:1999nt} and the too-big-to-fail problem~\cite{BoylanKolchin:2011de} remain. 

The observed rotation curves of baryonic matter in galaxies also motivates modified Newtonian dynamics (MOND) as an explanation for the problem \cite{Milgrom:1983ca}. In MOND the acceleration of test particles is modified, with respect to the Newtonian prediction, below a characteristic acceleration scale $a_0\sim c H_0$, where $c$ is the speed of light and $H_0$ the Hubble constant today. This modification accounts for the approximately flat asymptotic velocities of the galactic rotation curves at large radii \cite{Sanders:2002pf,Gentile:2010xt,McGaugh:2016leg,Lelli:2017vgz,Li:2018tdo} and the correlation of this asymptotic velocity with the total baryonic mass in the galaxy, i.e. the baryonic Tully-Fisher relation \cite{Tully:1977fu,McGaugh:2000sr}. On larger scales it has been found that MOND cannot account for the entire missing mass in galaxy clusters
\cite{Sanders:2002ue} or the dynamics of cluster mergers \cite{Angus:2006qy,Angus:2006ev}. Nor is it obvious if MOND can account for cosmological observations \cite{Skordis:2005xk,Dodelson:2006zt,Dodelson:2011qv}. For a recent review of the observational status of MOND see \cite{Famaey:2011kh}.

Here we study galactic rotation curve data and the predicted curves in $(g_{\rm  bar},g_{\rm tot})$-space ($g2$-space) from MOND and DM models with $g_{\rm tot}(r)$ being the total observed centripetal acceleration of matter in a rotationally supported galaxy as function of radial distance $r$ from the center. Similarly $g_{\rm  bar}(r)$ is the centripetal acceleration arising from the baryonic (visible) matter distribution assuming Newtonian gravity. 

We consider the predictions from two variants of MOND known as MOND modified inertia (MI) models \cite{Milgrom:1983ca,Milgrom:1992hr} which have been extensively employed to fit rotation curves \cite{Sanders:2002pf,Famaey:2005fd,Gentile:2010xt,McGaugh:2016leg,Lelli:2017vgz,Li:2018tdo}  and MOND modified gravity (MG) models in the Bekenstein-Milgrom formulation \cite{Bekenstein:1984tv}. In the latter case we employ an analytic approximation for the predicted rotation curves  \cite{Brada:1994pk}. For DM we consider the Navarro-Frenk-White \cite{Navarro:1996gj} and the quasi-isothermal density profiles. 

Rotation curve data from 175 galaxies in the SPARC database has recently been found to exhibit the distinct $g2$-space geometry of MOND modified inertia \cite{McGaugh:2016leg,Lelli:2017vgz,Li:2018tdo}, termed the Mass Discrepancy Acceleration Relation (MDAR) \cite{1990A&ARv21S,McGaugh:2004aw,2016MNRAS,McGaugh:2016leg} and this has motivated the study of models that mimick this behaviour, both in DM e.g. \cite{Chashchina:2016wle,Edmonds:2017zhg,Dai:2017unr,Cai:2017buj,Berezhiani:2017tth} and in modified gravity frameworks \cite{Burrage:2016yjm,Verlinde:2016toy,Vagnozzi:2017ilo}. It was however also found that ordinary Cold DM gives rise to this relation in the EAGLE simulation~\cite{Ludlow:2016qzh}.

In this study we find that MOND modified inertia, independent of the specific model used, is disfavoured by the data at more than 5$\sigma$. More generally this holds for any model yielding a monotonically increasing function in $g2$-space.

\bigskip
This paper is organized as follows: 

In section~\ref{Sec: geometries} we illustrate different $g2$-space geometries using a simple exponential disk model of the baryonic content of galaxies in Fig~\ref{Fig:MONDfunctions}. We give a global classification of geometries using the relative locations of $r_{\rm bar}$ and $r_{\rm tot}$ ---  
the radii of maximum baryonic and total accelerations respectively --- summarized in table~\ref{Table:geometries}. We then consider ratios of accelerations, $\hat{g}_{\rm bar}(r)\equiv g_{\rm bar}(r)/g_{\rm bar}(r_{\rm bar})$ and $\hat{g}_{\rm tot}(r) \equiv g_{\rm tot}(r)/g_{\rm  tot}(r_{\rm bar})$ and illustrate the $\hat{g}2$-space geometries in Fig.~\ref{Fig:MONDfunctionsscaled}. 

In section~\ref{Sec:data analysis} we present our analysis of the SPARC rotation curve data \cite{Lelli:2016zqa} using the full inferred baryonic matter distribution, including disk, bulge and gas components. The data is shown in $g2$-space and $\hat{g}2$-space in Fig.~\ref{Fig:Galaxies}. The latter eliminates systematic uncertainties on inclination angles and galaxy distances and reduces systematic uncertainties on mass-to-light ratios in the data. 

We first show that the prediction  $r_{\rm bar}=r_{\rm tot}$ from MOND modified inertia models, and consequently that $\hat{g}_{\rm bar,tot}(r_{\rm bar})=\hat{g}_{\rm bar,tot}(r_{\rm tot})$,  is in disagreement with data at more than 5$\sigma$. This is summarized in table~\ref{Table:datasets1}.

We next group the galaxies in SPARC according to the relative locations of $r_{\rm bar}$ and $r_{\rm tot}$, summarized in table~\ref{Table:data} and show the distribution of data in $\hat{g}2$-space at radii above and below $r_{\rm bar}$ for the full SPARC data set and for each of these groups in Fig.~\ref{Fig:datascaled}. The average $\hat{g}2$-space values of the full data set displays the characteristic geometry of DM with an isothermal density profile. This geometry is shared by the Bekenstein-Milgrom formulation of MOND modified gravity in the approximation used here. 
However the spread in data is significant. 
A minority of galaxies ---which by selection have data only at large radii - display the characteristic geometry of MOND modified inertia on average while another minority displays that of DM with an NFW profile.

In section~\ref{Sec:Summary} we summarize results and briefly discuss the limitations of our data analysis with respect to MOND modified gravity models and the relevance of improved measurements of rotation curves at small and moderate radii to probe the solution to the missing mass problem.

\section{Model geometries in $g2$-space }
\label{Sec: geometries} 
\noindent We begin by illustrating the geometry of MOND and DM models in $g2$-space in a simplified setting with the baryonic matter modelled purely as an infinitely thin disk with an exponential surface mass density 
\begin{equation}
\Sigma(r)=\Sigma_0 e^{-r/r_d} \ ,
\label{Eq:disk}
\end{equation}
where $\Sigma_0$ is the central surface mass density and $r_d$ is the scale length. 
For all quantitative results later we instead use the inferred baryonic accelerations from the SPARC database \cite{Lelli:2016zqa}.
We distinguish between two classes of MOND models that yield distinct geometries in $g2$-space, namely MOND modified inertia models (MI) \cite{Milgrom:1983ca,Milgrom:1992hr} --- in which the Newtonian equation of motion is modified but Newtonian gravity is not  --- and MOND modified gravity models (MG) in the formulation of Bekenstein-Milgrom \cite{Bekenstein:1984tv} in which the law of gravity itself is modified.  Below we will refer to the total centripetal acceleration of a test mass in the midplane of a disk galaxy, of an unspecified model, as $g_{\rm tot}$. The acceleration stemming from the visible matter assuming Newtonian gravity is termed $g_{\rm bar}$. Finally when discussing specific models we will refer to the total acceleration with subscripts corresponding to that model, like $g_{\rm MI}$ for the total acceleration in a MOND modified inertia model.  

\bigskip
{\bf MOND Models:} 
In MOND modified inertia models the total centripetal acceleration, $g_{\rm MI}$, on a test mass in the galactic plane is related to the Newtonian one, $g_{\rm bar}$, via the relations 
\begin{equation}
 g_{\rm bar}(g_{\rm  MI})=\mu(x)g_{\rm MI} , \quad x\equiv \frac{g_{MI}}{g_0} ; \qquad \qquad g_{\rm MI}(g_{\rm  bar})=\nu(y)g_{\rm bar} , \quad y\equiv \frac{g_{bar}}{g_0} , 
\label{Eq: MONDI}
\end{equation}
where  $g_0\sim 10^{-10} \frac{m}{s^2}$ is the characteristic acceleration scale of MOND. The interpolation function $\mu(x)$ smoothly interpolates between the deep Mondian regime $\mu(x)\simeq x$ for $x\ll 1$ and the Newtonian regime $\mu(x)\simeq 1$ for $x\gg 1$, but is otherwise undetermined at this level where a complete model of MOND modified inertia is not specified. 
The inverse interpolation function is $\nu(y)\equiv I^{-1}(y)/y$ with $I(x)=x \mu(x) =y$. Consequently in MOND modified inertia $g_{\rm  MI}( g_{\rm bar})$ is a single valued function of  $g_{\rm bar}$.

In the Bekenstein-Milgrom formulation of MOND modified gravity models \cite{Bekenstein:1984tv} the total centripetal acceleration is determined via a modified Poisson equation for the MOND potential field $\psi$
\begin{equation}
\vec{\nabla} \cdotp (\mu(\frac{| \vec{\nabla}\psi |}{g_0}) \vec{\nabla} \psi  ) = 4 \pi G \rho , 
\end{equation}
where the properties of the undetermined interpolation function is as above for MOND modified inertia. By noting that $4 \pi G \rho=  \vec{\nabla} \cdot \vec{g}_{\rm bar}$, solutions to this equation are of the form 
\begin{equation}
 \vec{g}_{\rm bar}=\mu(\frac{| \vec{\nabla}\psi |}{g_0})\vec{\nabla} \psi +\vec{\nabla} \times \vec{ h} \ ,
\label{eq3}
\end{equation}
where $h$ is a generic vector field.  
An approximate expression for the resulting acceleration $g_{\rm MG}$ in MOND modified gravity, analogous to that in Eq.~\ref{Eq: MONDI}, for an exponential disk galaxy is derived in \cite{Brada:1994pk}: 
\begin{align}
 g_{\rm bar}(g_{MG}, r) &= \mu(\frac{g_{\rm MG}^+}{g_0})  g_{\rm MG} , \quad   g_{\rm MG}(g_{\rm bar}, r)= \nu(\frac{g_{\rm bar}^+}{g_0})  g_{\rm bar} ; 
 \\
 \quad   g_{\rm MG}^+ &=I^{-1}(g_{\rm bar}^+) ,\quad  g_{\rm bar}^+(g_{\rm bar},r)=\sqrt{g_{\rm bar}^2+ (2 \pi G \Sigma(r))^2} . 
 \label{Eq:MONDgrav}
\end{align}
Due to the radial dependence of the fiducial quantities $g_{{\rm bar},MG}^+$ the MOND modified gravity acceleration $g_{\rm  MG}( g_{\rm bar},r)$ is not a single valued function of the baryonic acceleration $g_{\rm bar}$.

A number of interpolation functions $\mu(x)$ and inverse interpolation functions $\nu(y)$ have been considered in the literature, e.g. \cite{Begeman:1991iy,Bekenstein:2004ne}.  
For our analysis the details of the interpolation function are not central and we therefore focus on the inverse interpolation function from \cite{McGaugh:2008nc,Famaey:2011kh} which was used to fit the SPARC galaxy data in \cite{McGaugh:2016leg,Lelli:2017vgz}: 
\begin{equation}
\nu(y)=\frac{1}{1-e^{-\sqrt{y}}} . 
\label{Eq:nufunction}
\end{equation}
\normalsize

\begin{figure}[htp!]
	\centering
	\includegraphics[width=0.3\textwidth]{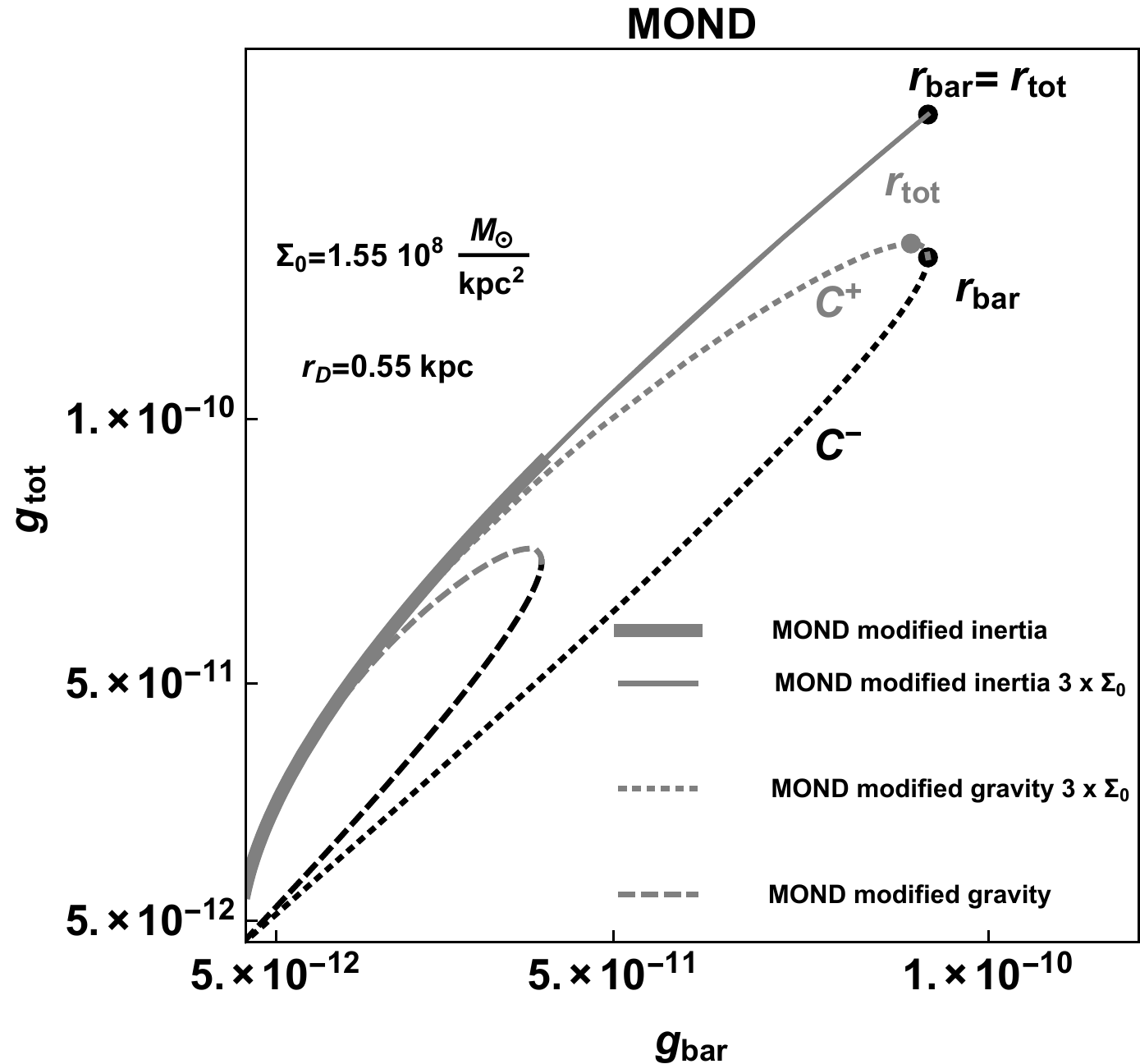}
	\includegraphics[width=0.32\textwidth]{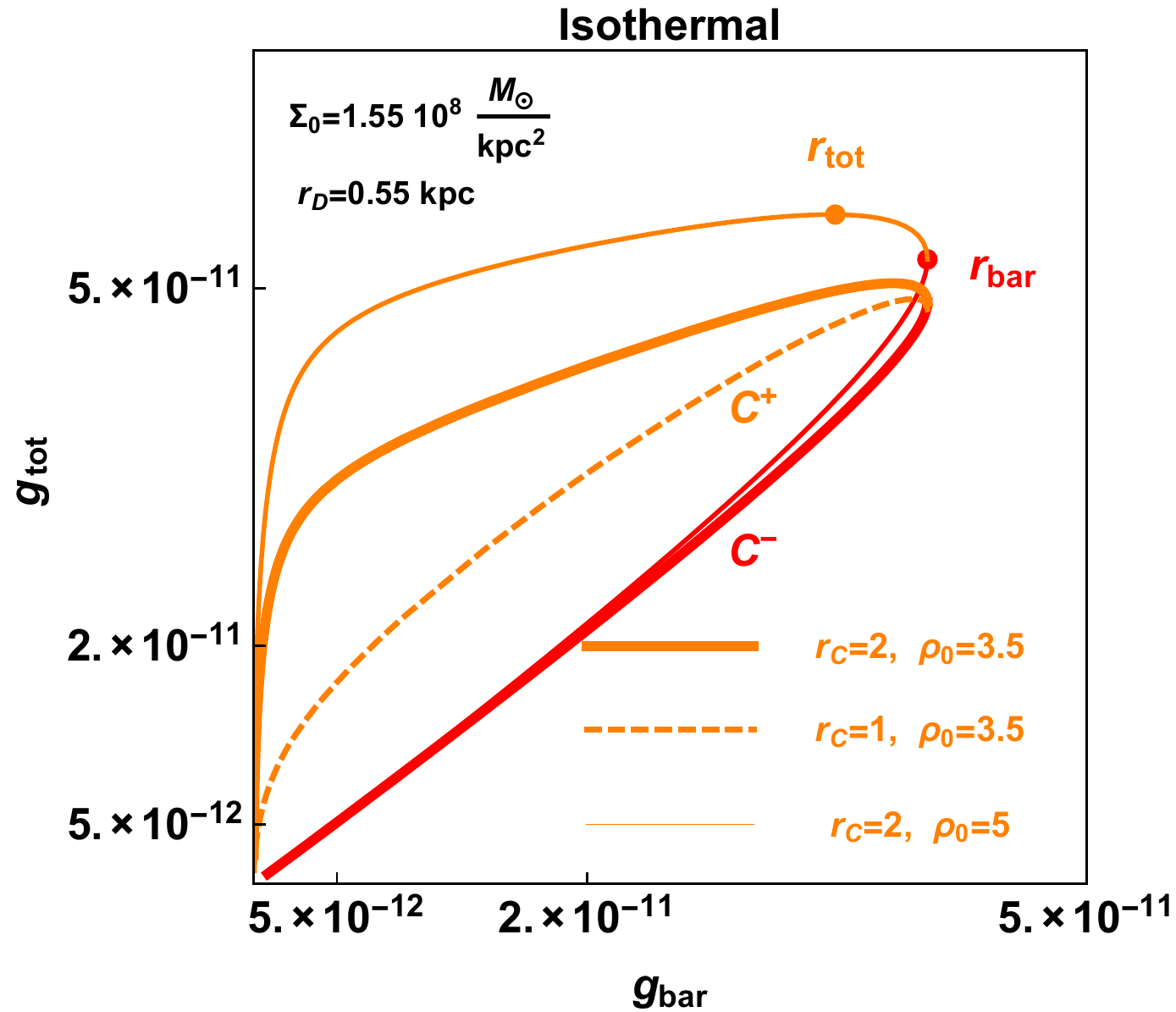}
\includegraphics[width=0.32\textwidth]{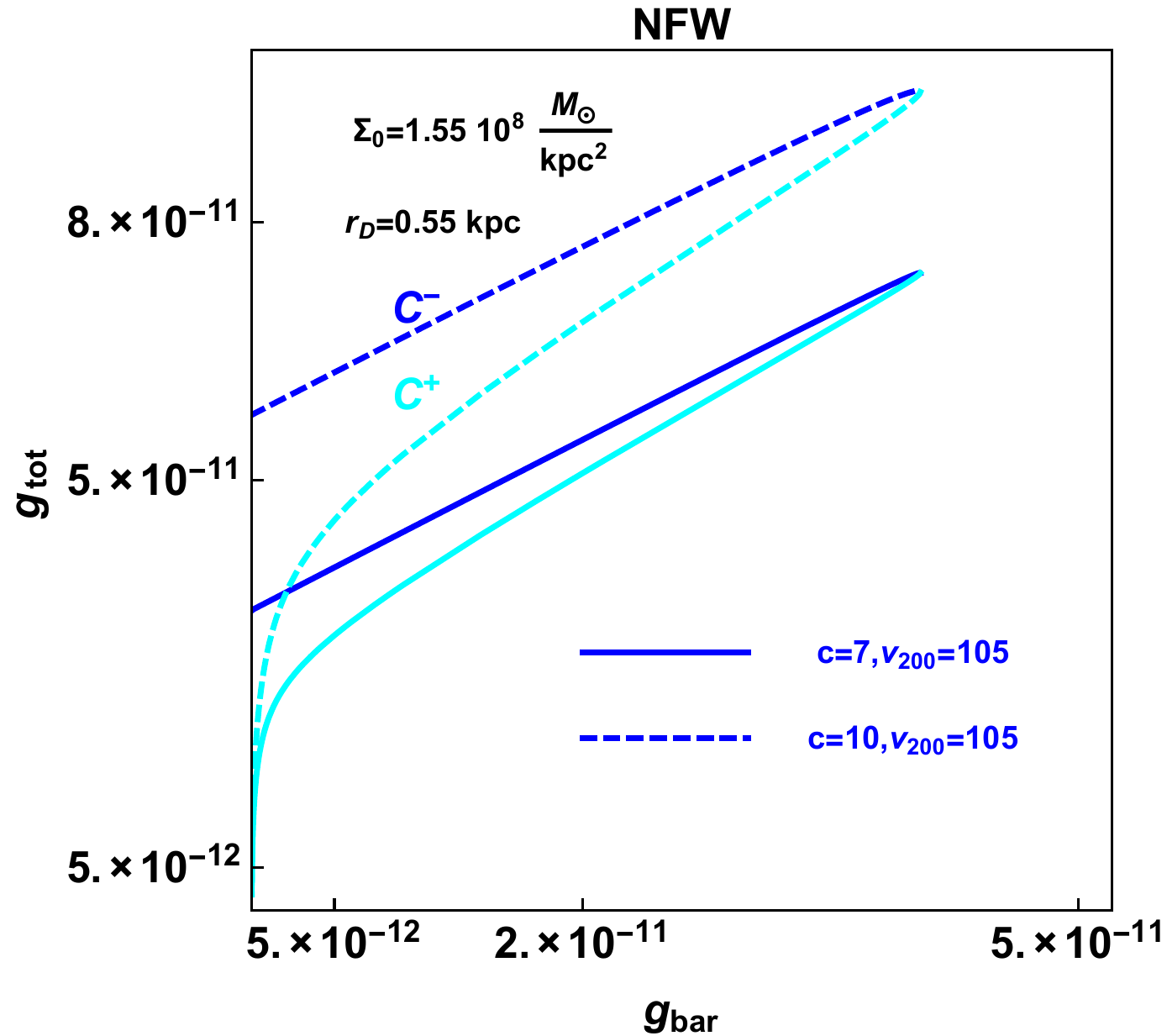}
	\caption
	{The $g2$-space geometry of MOND and DM models using texponential disk in Eq.~\ref{Eq:disk} for the baryonic matter with scale height  $r_D=0.55$ kpc.
	\newline
	{\it Left Panel}:  MOND modified inertia (solid thick and thin) curves and MOND modified gravity (dotted and dashed curves) with interpolation function in Eq.~(\ref{Eq:nufunction}), and for MOND modified gravity, the approximation in Eq.~\eqref{Eq:MONDgrav}. We show two values of the central surface density $\Sigma_0=1.55\times 10^8 M_\odot/\rm{kpc}^2$ (solid thin and dotted) and  $3 \times \Sigma_0$ (solid thick and dashed). 
	The reference radii $r_{\rm bar}$ and $r_{\rm tot}$ defined in Eq.~\ref{Eq:refradii} and the curve segments $\mathcal{C}^{\pm}$ as defined in Eq.~\ref{Eq:refcurves} are shown for two of the curves. \newline
	{\it Middle panel}: DM models with quasi-isothermal density profiles for different values of the DM density scale radius $r_C$ and central density $\rho_0$.\newline
	{\it Right Panel}: DM models with NFW density profiles for different values of velocity $v_{200}$ at the virial radius $r_{200}$ and the concentration parameter $c=r_{200}/r_s$ and $v_{200}$.
	}
	\label{Fig:MONDfunctions}
\end{figure}

In order to classify $g2$-space geometries and rotation curve data we define two reference radii, $r_{\rm bar}$ and $r_{\rm tot}$ as the radii at which $g_{\rm bar}$ and $g_{\rm tot}$ are maximum respectively, 
\begin{equation}
g_{\rm bar}(r_{\rm bar}) = {\rm max} \{ g_{\rm bar}(r) \} , \quad g_{\rm tot}(r_{\rm tot}) = {\rm max} \{ g_{\rm tot}(r) \}  \ . 
\label{Eq:refradii}
\end{equation}
We also define the curve segments $\mathcal{C}^{\pm}$ above and below $r_{\rm bar}$ (similarly we could use $r_{\rm tot}$ as reference radius) of a given model in $g2$-space as
\begin{equation}
\mathcal{C}^{\pm}= \{ (g_{\rm bar}(r),  g_{\rm tot}(r)) ; r \gtrless  r_{\rm bar}\} . 
\label{Eq:refcurves}
\end{equation}
In the left panel of Fig.~\ref{Fig:MONDfunctions} we show the MOND modified inertia curve 
from Eq.~(\ref{Eq: MONDI}) (solid line) and the approximate modified gravity curve from 
 Eq.~(\ref{Eq:MONDgrav}) (dotted and dashed curves). 
The reference radii $r_{\rm bar,obs}$ are indicated with dots while the grey curve segments correspond to $\mathcal{C}^+$ and the black curve segments to $\mathcal{C}^-$.  

In MOND modified inertia models $r_{\rm bar}=r_{\rm tot}$ and the two curve segments coincide, ie. $\mathcal{C}^-=\mathcal{C}^+$, as consequences of the MOND modified inertia function $g_{\rm MI}(g_{\rm bar})$ being single valued. Equivalently, the area enclosed by the MOND modified inertia curve $\mathcal{C}_{\rm MI}$ is zero, $\mathcal{A}(\mathcal{C}_{\rm MI})=0$ as discussed in~\cite{Petersen:2017klw}. 

In the MOND modified gravity approximation of Eq.~\eqref{Eq:MONDgrav} it follows that $r_{\rm bar}<r_{\rm tot}$ and the curve segment $\mathcal{C}^+$ is above the curve segment  $\mathcal{C}^-$ in $g2$-space. Equivalently, the enclosed area of the MOND modified gravity curve is non-zero $\mathcal{A}(C_{\rm MG})>0$. We summarize these properties in the first two rows of table~\ref{Table:geometries}.

In  Fig.~\ref{Fig:MONDfunctions} we have used the exponential disk in Eq.~(\ref{Eq:disk}) for the baryonic matter distribution, for which $r_{\rm bar}\simeq 0.41 r_d$ and the interpolation function corresponding to Eq.~\eqref{Eq:nufunction}.  
Since $g_{\rm bar}(r=0)=g_{\rm bar}(r=\infty)=0$ the curves shown are closed with curve parameters $0 \leq r \leq \infty$. 
The scale length $r_d$ of the exponential disk does not influence the geometry of the curves but only how much of the curve is traced up to a given radius $r$. 
The central surface density $\Sigma_0$ scales the maximum values of $g_{\rm bar}$, and $g_{\rm tot}$ and therefore stretches or shrinks the curves. For MOND modified inertia, curves with smaller $\Sigma_0$ coincide with a part of those with a larger $\Sigma_0$. For MOND modified gravity we illustrate the shrinking and stretching by plotting two different values of $\Sigma_0$ \footnote{If the galactic mass is kept fixed $r_d$ and $\Sigma_0$ cannot be varied independently}.
 In both cases,  for a given interpolation function and acceleration scale $g_0$, the $g2$-space curves are completely determined for all galaxies by the baryonic matter distribution.
. 

\bigskip
{\bf Dark Matter:} 
In DM models the total centripetal acceleration $g_{\rm DM}(r)=g_{\rm bar}(r)+g_{\rm halo}(r)$ is a sum of the contributions from the baryonic and DM density distributions --- here assumed to be a spherical halo for simplicity.
To illustrate the $g2$-space geometry of the considered dark matter models we again employ the exponential disk in Eq.\eqref{Eq:disk} for the baryonic matter and two different DM density profiles 
\begin{equation}
\rho_{\rm NFW}(r)=\frac{\rho_{0,{\rm NFW}}}{\frac{r}{r_s}(1+\frac{r}{r_s})^2} , \quad \rho_{\rm ISO}(r)=\frac{\rho_{0,{\rm ISO}}}{1+(\frac{r}{r_c})^2} \ ,
\end{equation}
where $\rho_{0,{\rm NFW}},\rho_{0,{\rm ISO}}$ are mass densities and $r_s,r_c$ are scale lenghts respectively. 
The Navarro-Frenk-White profile $\rho_{\rm NFW}(r)$ is motivated by fits to the density of halos in simulations of cold collisioness DM \cite{Navarro:1995iw} and leads to a cuspy central DM density profile at small radii scaling as $\rho_{\rm NFW}(r)\sim r^{-1}$. The quasi-isothermal DM density profile $\rho_{\rm ISO}(r)$ may be physically realized (at small radii) in models with sizeable DM self interactions and leads to a cored DM density profile at small radii scaling as $\rho_{\rm ISO}(r)\sim r^{0}$. 
It has recently been proposed that the diversity of galactic rotation curves \cite{Navarro:2008kc} can be accomodated in a model of self interacting DM where the resulting DM density profile is approximately quasi-isothermal profile at small radii, set by the DM density and self-interaction cross-section, while following the NFW profile at large radii \cite{Kamada:2016euw,Creasey:2016jaq}. For both density profiles the centripetal accelerations in the midplane of a disk galaxy $g_{NFW}(g_{\rm bar},r)$, $g_{ISO}(g_{\rm bar},r)$ are not single valued functions of $g_{\rm bar}$.

We show examples of DM model curves in $g2$-space for the quasi-isothermal and NFW profiles respectively in the middle and right panels of Fig.~\ref{Fig:MONDfunctions}. The curve segments  $\mathcal{C}^+$ are shown in orange and cyan respectively while the curve segments  $\mathcal{C}^-$ are shown in red and blue respectively.
The full curves in the quasi-isothermal case are closed curves, since also $g_{\rm ISO}(r=0)=g_{\rm ISO}(\infty)=0$ while the area of the curve is non-zero $\mathcal{A}(C_{\rm ISO})>0$ as discussed in~\cite{Petersen:2017klw}.   
The width of the curve is controlled by $\rho_0$, as seen by comparing the solid thick 
and solid thin 
curves, while the steepness of the curve near $r=0$ is controlled by $r_c$ as seen by comparing the dashed and dotted curves. 
. 

The NFW curves are distinct by not being closed due to the divergence of the profile at small $r$ --- the cuspyness of the NFW profile translates into $g_{\rm tot, DM}(r=0)> 0$ --- and by the fact that the curve segments $\mathcal{C}^+$ lie below the curve segments $\mathcal{C}^-$. The width of the NFW curve is controlled by the concentration parameter $c=\frac{r_{\rm vir}}{r_s}$, where $r_{\rm vir}$ is the virial radius, as seen by comparing the solid and dashed curves.

We summarize the characteristics of the model geometries in Table~\ref{Table:geometries}. For MOND modified inertia models $r_{\rm tot}=r_{\rm bar}$ and $\mathcal{C}^-=\mathcal{C}^+$ and consequently $\mathcal{A}(\mathcal{C}_{\rm MI})=0$. For MOND modified gravity and quasi-isothermal DM models $r_{\rm tot}>r_{\rm bar}$ and the curve segments $\mathcal{C}^+$ lie above $\mathcal{C}^-$ in $g_{\rm tot}$ values and consequently $\mathcal{A}(\mathcal{C}_{\rm ISO})>0$. Finally for NFW DM models the curve segments $\mathcal{C}^-$ lie above $\mathcal{C}^+$ in $g_{\rm tot}$ (with $r_{\rm tot}<r_{\rm bar}$ barely visible) and the area is undefined.
The degeneracy of the MOND modified gravity approximation and DM-ISO geometries with respect to these basic characteristics, does not imply the geometry is identical as is evident from  Fig.~\ref{Fig:MONDfunctions}. In particular the shape of the DM-ISO curves is controlled by the scale length of the DM density an additional free parameter as compared to the MOND modified gravity approximation.

  \bigskip
 \begin{table}[t!]
\centering
\begin{tabular}{ |p{2.5cm}||p{3cm}|p{3cm}|p{3cm}| }
 \hline
Models& Reference radii & Curve segments& Curve Area\footnote{The curves are closed and the areas spanned by the curves, $\mathcal{A}(\mathcal{C})$, are defined for the first three models provided the baryonic accelerations satisfy $g_{\rm bar}(r=0)= g_{\rm bar}(r=\infty)=0$ as is the case for an exponential disk.}\\
 \hline
MOND-MI &   $r_{\rm tot}=r_{\rm bar}$ & $\mathcal{C}^+=\mathcal{C}^-$&$\mathcal{A}(\mathcal{C})=0$\\
MOND-MG &   $r_{\rm tot}>r_{\rm bar}$ &$\mathcal{C}^+>\mathcal{C}^-$&$\mathcal{A}(\mathcal{C})>0$\\
\textcolor{red}{DM-ISO}  & \textcolor{red}{$r_{\rm tot}>r_{\rm bar}$ }&\textcolor{red}{$\mathcal{C}^+>\mathcal{C}^-$}&\textcolor{red}{$\mathcal{A}(\mathcal{C})>0$}\\
\textcolor{blue}{DM-NFW}   &\textcolor{blue}{$r_{\rm tot}<r_{\rm bar}$}& \textcolor{blue}{$\mathcal{C}^+< \mathcal{C}^-$} &\textcolor{blue}{ Curves open} \\
 \hline
\end{tabular}
\caption{Global characteristics of geometries in $g2$-space for MOND and DM models as shown in Fig.~\ref{Fig:MONDfunctions}. The reference radii $r_{\rm tot}$ and $r_{\rm bar}$ are the radii of maximum total acceleration and maximum inferred baryonic acceleration, as defined in Eq.~\eqref{Eq:refradii} and the curve segments $\mathcal{C}^{\pm}$ are defined in Eq.~\eqref{Eq:refcurves}.
}
\label{Table:geometries}
\end{table}
\bigskip

\begin{figure}[htp!]
	\centering
	\includegraphics[width=0.32\textwidth]{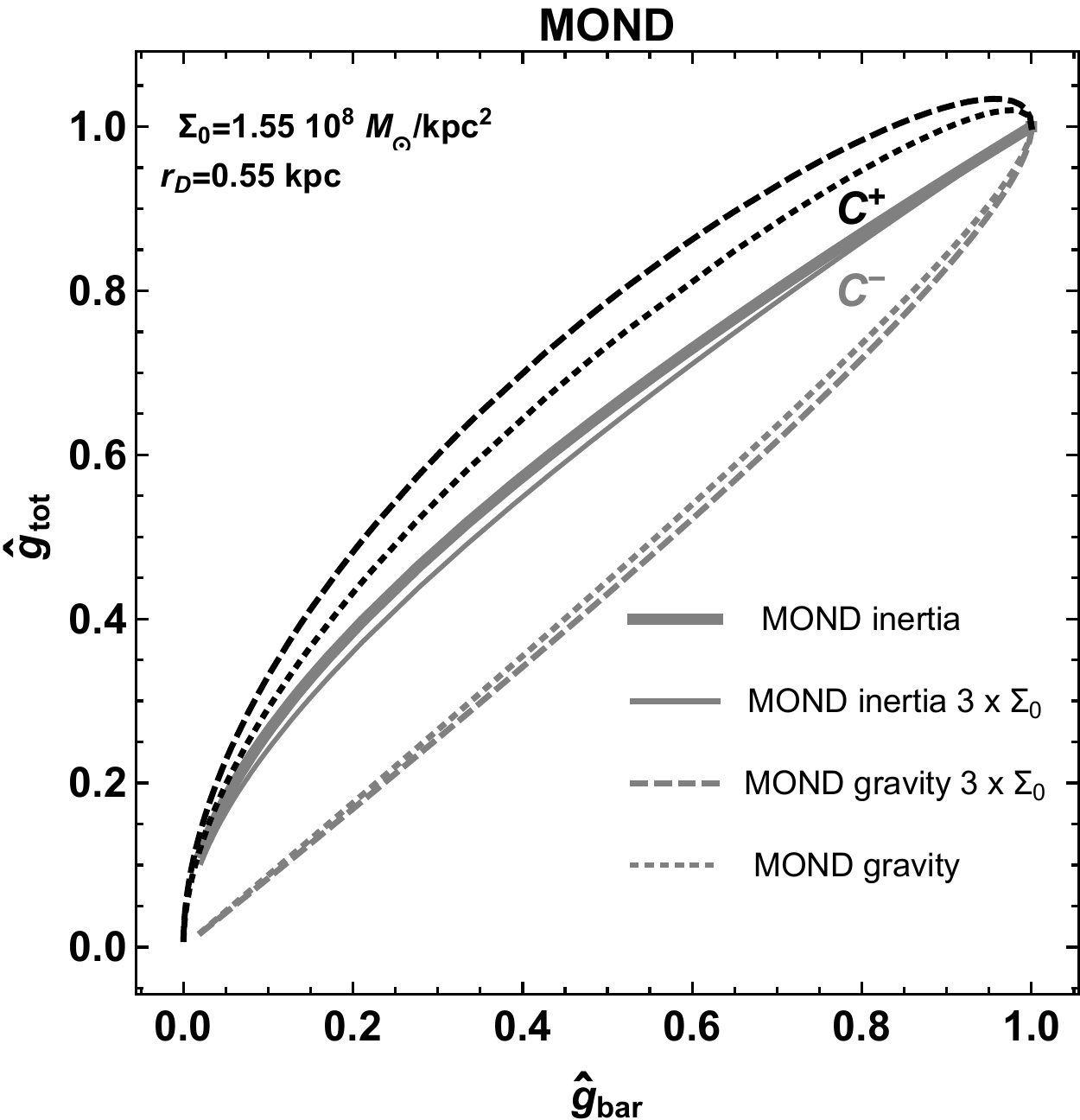}
	\includegraphics[width=0.32\textwidth]{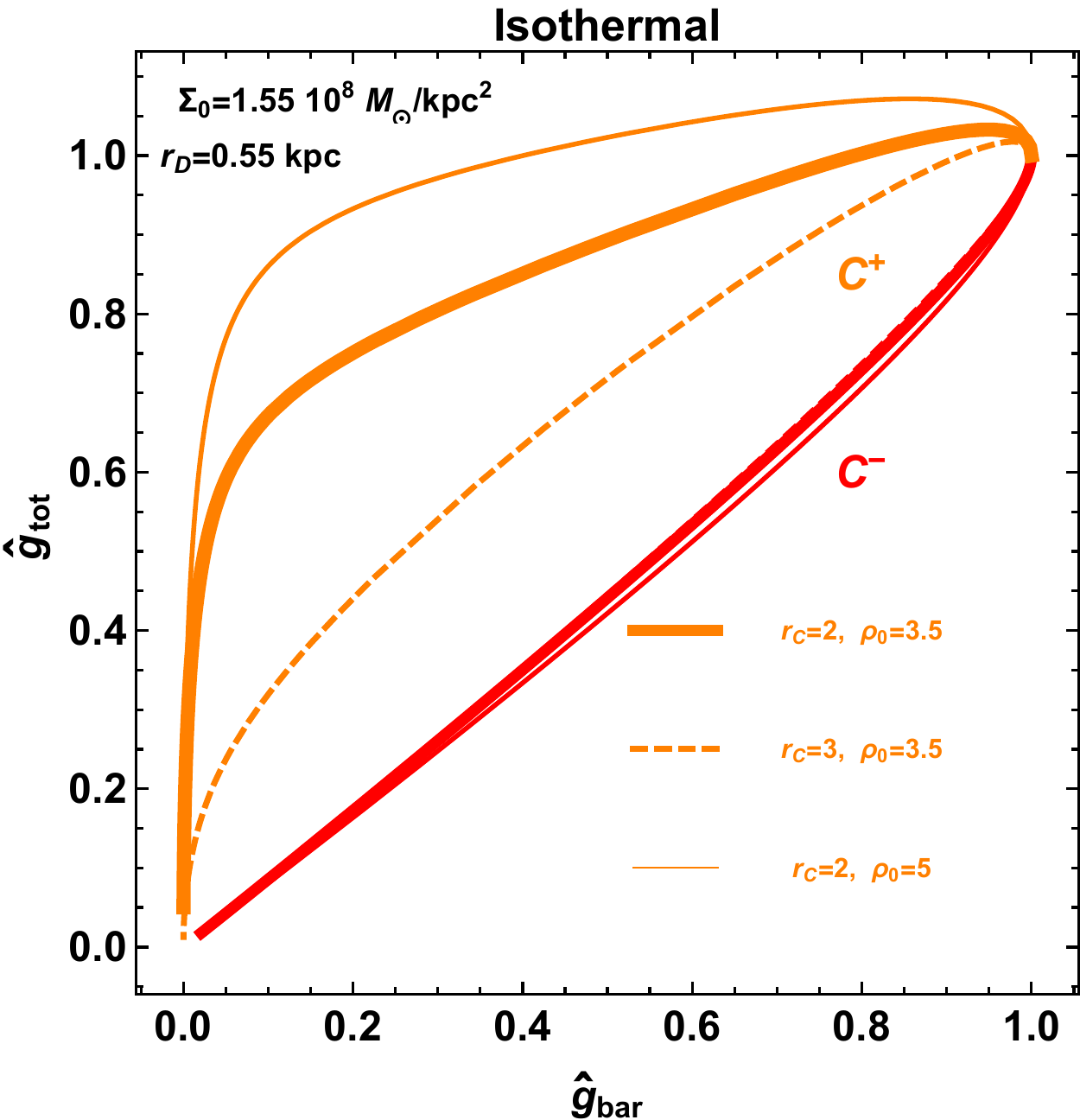}
	\includegraphics[width=0.32\textwidth]{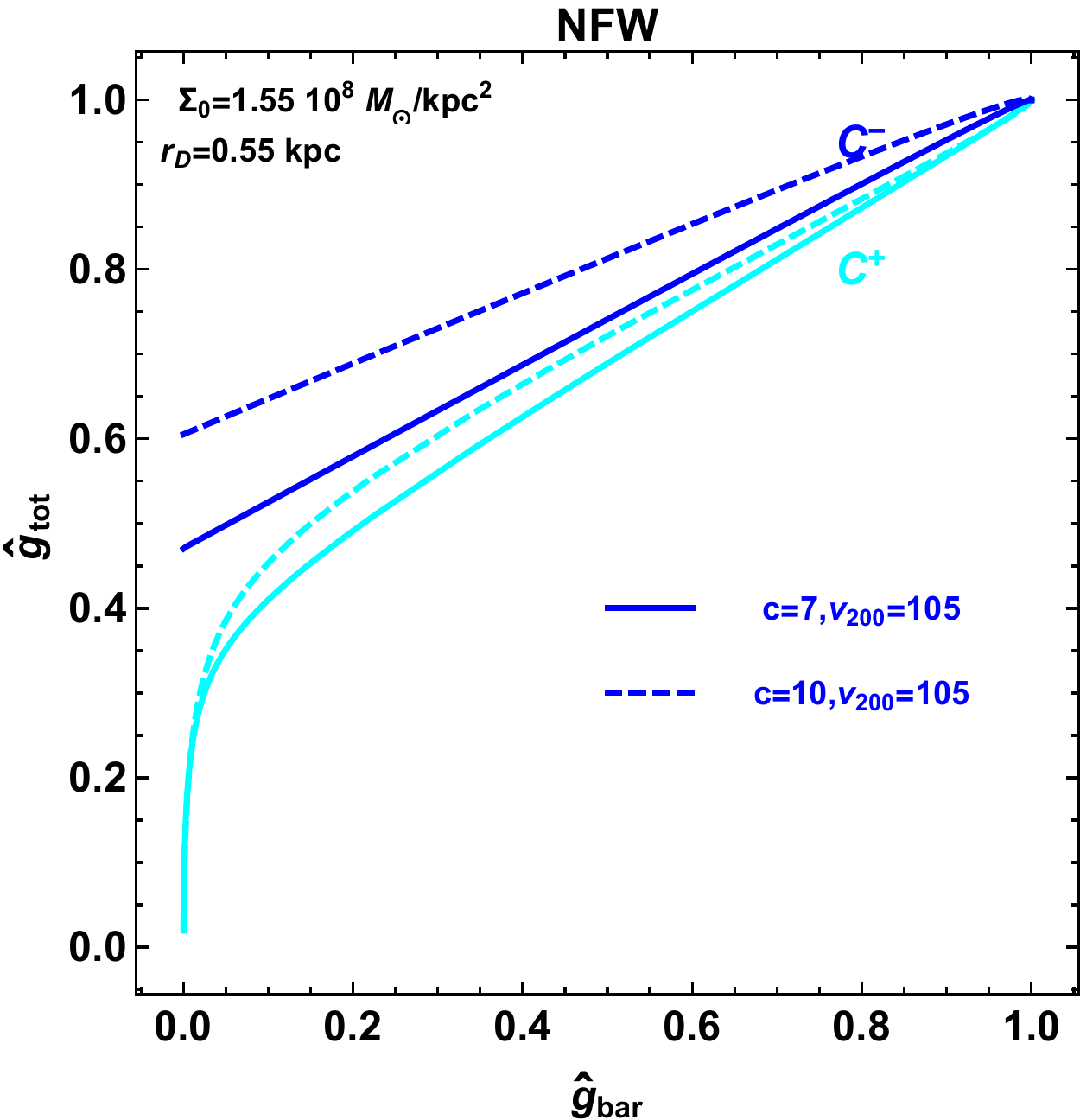}
	\caption
	{The $\hat{g}2$-space geometry $(\hat{g}_{\rm bar}(r),\hat{g}_{\rm tot}(r))$ of MOND and DM models as defined in Eq.~(\ref{Eq:normg2-space}). Curves and parameters are otherwise identical to those in Fig.~\ref{Fig:MONDfunctions}.}
	\label{Fig:MONDfunctionsscaled}
\end{figure}

{\bf Normalized $\hat{g}2$-space:} 
In order to display the average geometry of several galactic rotation curves and to reduce systematic uncertainties it is relevant to consider ratios of accelerations 
in a normalized $\hat{g}2$-space by defining
\begin{align}
\hat{g}_{\rm bar, tot}(r) \equiv g_{\rm bar, tot}(r)/g_{\rm bar, tot}(r_{\rm bar}) .
\label{Eq:normg2-space}
\end{align}
Another possibility here would be to use $r_{\rm tot}$ as a reference radii in the denominator above. 
 We replot the MOND and DM model geometries from Fig.~\ref{Fig:MONDfunctions} in the rescaled $\hat{g}2$-space in Fig~(\ref{Fig:MONDfunctionsscaled}).

\section{data analysis}
\label{Sec:data analysis}
\noindent We study rotation curve data from the 175 galaxies in the SPARC database \cite{Lelli:2016zqa}. The database provides the observed total rotational velocities $v_{\rm obs}(r_j)$, as a function of observed radii points $r_j$.
The database also provides the inferred rotational velocities $v_{\rm disk}(r_j)$, $v_{\rm bul}(r_j)$, $v_{\rm gas}(r_j)$ from the baryonic matter components of the galaxies, divided into stellar disks, bulges and gas components. From this we compute the inferred baryonic acceleration $g_{\rm bar}(r_j),$ and the total observed acceleration $g_{\rm obs}(r_j)$ at each radii $r_j$ as
\begin{equation}
 g_{\rm obs}(r_j) = \dfrac{v_{\rm obs}^{2}(r_j)}{r_j},  \quad g_{\rm bar}(r_j) = \frac{\left(v_{\rm gas}^2(r_j) + \Upsilon_{\rm disk}v_{\rm disk}^2(r_j) + \Upsilon_{\rm bul}v_{\rm bul}^2(r_j)\right)}{r_j}  .
\end{equation}
We adopt as central values for the mass to light ratios $\Upsilon_{\rm disk}= 0.5\frac{M_{\odot}}{L_{\odot}}$ and $\Upsilon_{\rm bulge}= 0.7\frac{M_{\odot}}{L_{\odot}}$.
The SPARC data base also provides the corresponding (random) uncertainties $\delta {v_{\rm obs}}(r_j)$, as well as the uncertainties $\delta i $ and $\delta D$ on the galaxy inclination angle $i$ and distance $D$.  Following \cite{Lelli:2016zqa} we further adopt a 10 percent uncertainty on $v_{\rm gas}$ and 25 percent uncertainties on $\Upsilon_{\rm disk,bulge}$, i.e. $\delta {v_{\rm gas}}=0.1 v_{\rm gas}$ and 
 $\delta \Upsilon_{\rm disk,bulge}= 0.25 \Upsilon_{\rm disk,bulge}$. With this input we compute the $  \delta {g_{\rm bar}},  \delta {g_{\rm obs}} $ uncertainties
 \begin{align}
 \delta {g_{\rm obs}}(r_j)  &= g_{\rm obs}(r_j)  
 \sqrt{\bigg[\dfrac{2\delta{v_{\rm obs}} (r_j) }{v_{\rm obs}(r_j) }\bigg]^{2} 
 + \bigg[\dfrac{2\delta i}{\tan(i)}\bigg]^{2} 
 + \bigg[\dfrac{\delta D}{D}\bigg]^{2}},
 \\ \nonumber
  \delta {g_{\rm bar}}(r_j)  &= \frac{\sqrt{(2v_{\rm gas}(r_j) )^2 \delta v_{\rm gas}^2 + v_{\rm disk}^4(r_j) \delta \Upsilon_{\rm disk}^2 +v_{\rm bulge}^4(r_j) \delta \Upsilon_{\rm bulge}^2 } }{r_j} .
\label{Eq:uncertaintiest}
\end{align}
where we note that the inferred $g_{bar}(r_j)$ are independent of distance $D$ and inclination angle $i$ \cite{Li:2018tdo}.

We treat the uncertainties $\delta{v_{\rm obs}}, \delta{v_{\rm gas}}$  as random gaussian errors for each data point while the remaining uncertainties, $\delta i $, $\delta D$, $\delta \Upsilon_{\rm disk,bulge}$ are systematic errors, rescaling all data points within a galaxy in the same direction.  To reduce these systematic uncertainties we will  analyze ratios of accelerations as in Eq.~\eqref{Eq:normg2-space}, defining:
\begin{align}
\hat{g}(r_j)_{\rm bar, obs}=\frac{g(r_j)_{\rm bar, obs}}{g(r_{\rm bar })_{\rm bar, obs}} \ , \quad \hat{g}(\Delta r)_{\rm bar, obs}=\frac{g( \Delta r)_{\rm bar, obs}}{g( \Delta r_{\rm bar })_{\rm bar, obs}} \ ; \quad   
g( \Delta r)_{\rm bar, obs} \equiv \frac{1}{N_\Delta} \sum_{j \in \Delta  r} g(r_j)_{\rm bar, obs}  
\end{align}
where $\Delta r$ denotes an interval centered on $r$ that we average $g$ over within a galaxy,  $\Delta r_{\rm bar }$ is an equivalent interval around $r_{\rm bar }$ and $N_\Delta$ denote the number of points in the interval.

The ratios $\hat{g}_{\rm obs} (r_j)$ and $\hat{g}_{\rm obs} (\Delta r)$  eliminate the systematic uncertanties $\delta i , \delta D$ in galaxy inclination angle $i$ and galaxy distance $D$, up to any significant variation of inclination angle with radius within a single galaxy \cite{Li:2018tdo}, while $\hat{g}_{\rm obs} (\Delta r)$, reduces the systematic error introduced by the single normalization point $g_{\rm obs}(r_{\rm bar })$ in  $\hat{g}_{\rm obs} (r_j)$ when averaging over several galaxies. 
As we show explicitly in the appendix
$\hat{g}_{\rm bar}(r_j)$ and $\hat{g}_{\rm bar} (\Delta r)$  reduce the systematic uncertainties in $\delta \Upsilon_i$ significantly, especially near $r_{\rm bar}$ by construction,
where we are particularly interested in the geometry. 
These three sources of systematic uncertainties were found to be the dominant sources of scatter in previous analysis \cite{Li:2018tdo}. 
With the above construction there is only a small remaining systematic error on $\hat{g}_{\rm bar}(r_j)$ from mass to light ratios contained in the small quantity $\Delta \Upsilon$, in Eq.~\eqref{Eq:haterrors}. This means we can to a good approximation take the errors of $\hat{g}_{\rm obs,bar}(r_j)$ and $\hat{g}_{\rm obs,bar}(\Delta r)$ from different galaxies to be uncorrelated, even if the error on the mass to light ratios should be correlated for different galaxies.  
There is also a possible systematic uncertainty on $\hat{g}_{\rm obs} (\Delta r)$ from data points which may be included in both numerator and denominator when $\Delta r$ and $\Delta r_{\rm bar}$ overlap. This part of the error budget for $\hat{g}_{\rm obs} (\Delta r)$ is however completely uncorrelated between different galaxies under the assumption that $v_{\rm obs}$ values are uncorrelated. 
The details of the errors are discussed in the appendix~\ref{sec:app}.

\subsection{Data Selection}
\noindent We begin with the 175 galaxies in the SPARC database and discard $22$ galaxies based on the same quality criteria applied in \cite{McGaugh:2016leg,Lelli:2017vgz}. 
Ten of these are face-on galaxies with inclination angle $i < 30^0$ that are
rejected to minimize corrections to the observed velocities and 
twelve are galaxies with asymmetric rotation curves
that do not trace the equilibrium gravitational
potential. We discard one more galaxy, UGC01281, with large negative inferred speeds $v_{\rm gas}$ for the gas component leaving 152 galaxies with 3143 data points. A further data requirement $\delta {v_{\rm obs}}/v_{\rm obs}<0.1$ was imposed in \cite{McGaugh:2016leg,Lelli:2017vgz}. We only include this additional requirement when explicitly stated, e.g in the data sample $N_{G_2}$ discussed below, and otherwise keep all the 3143 data points. 

\begin{figure}[htp!]
	\centering
	\includegraphics[width=0.4\textwidth]{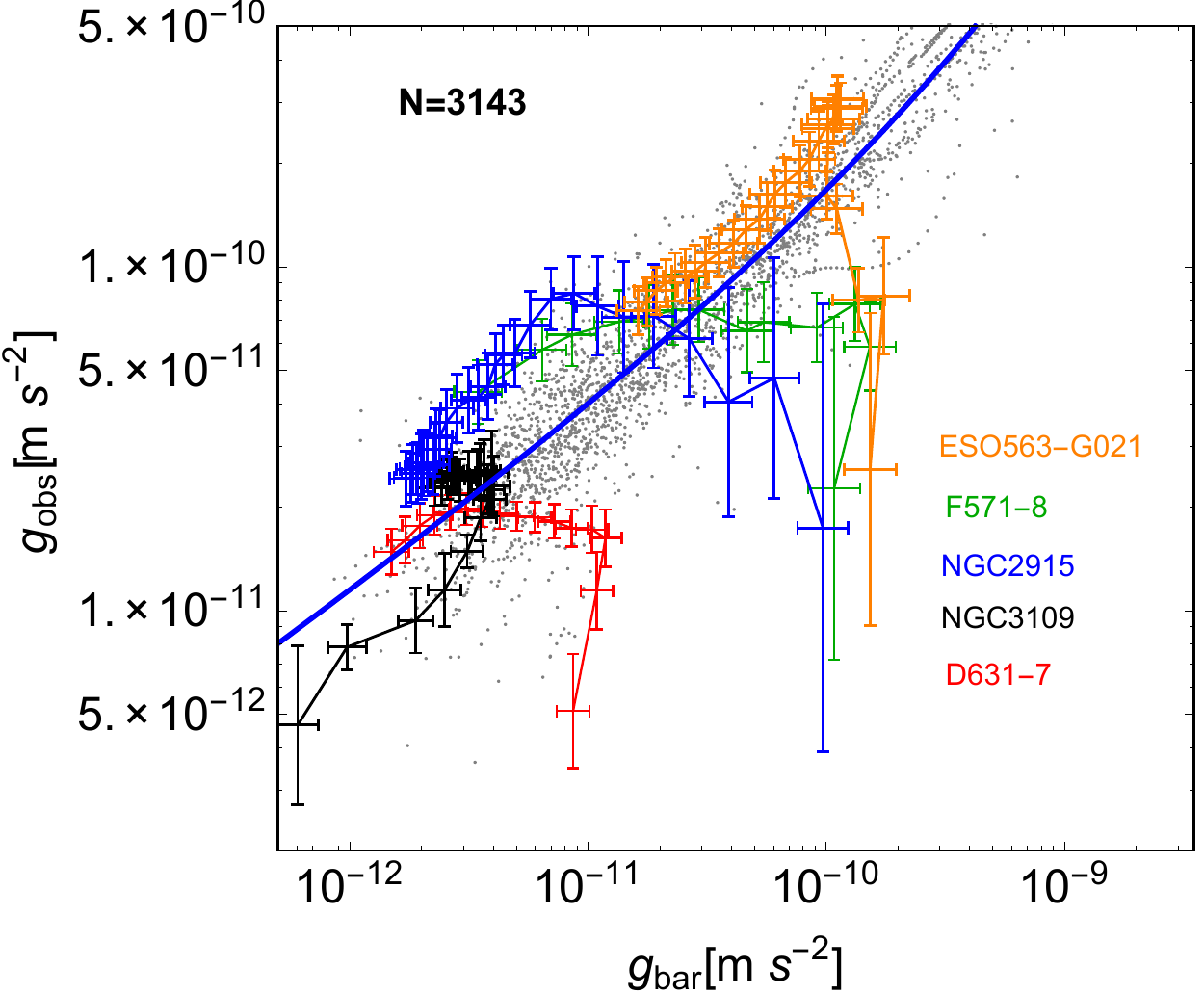}
	\includegraphics[width=0.4\textwidth]{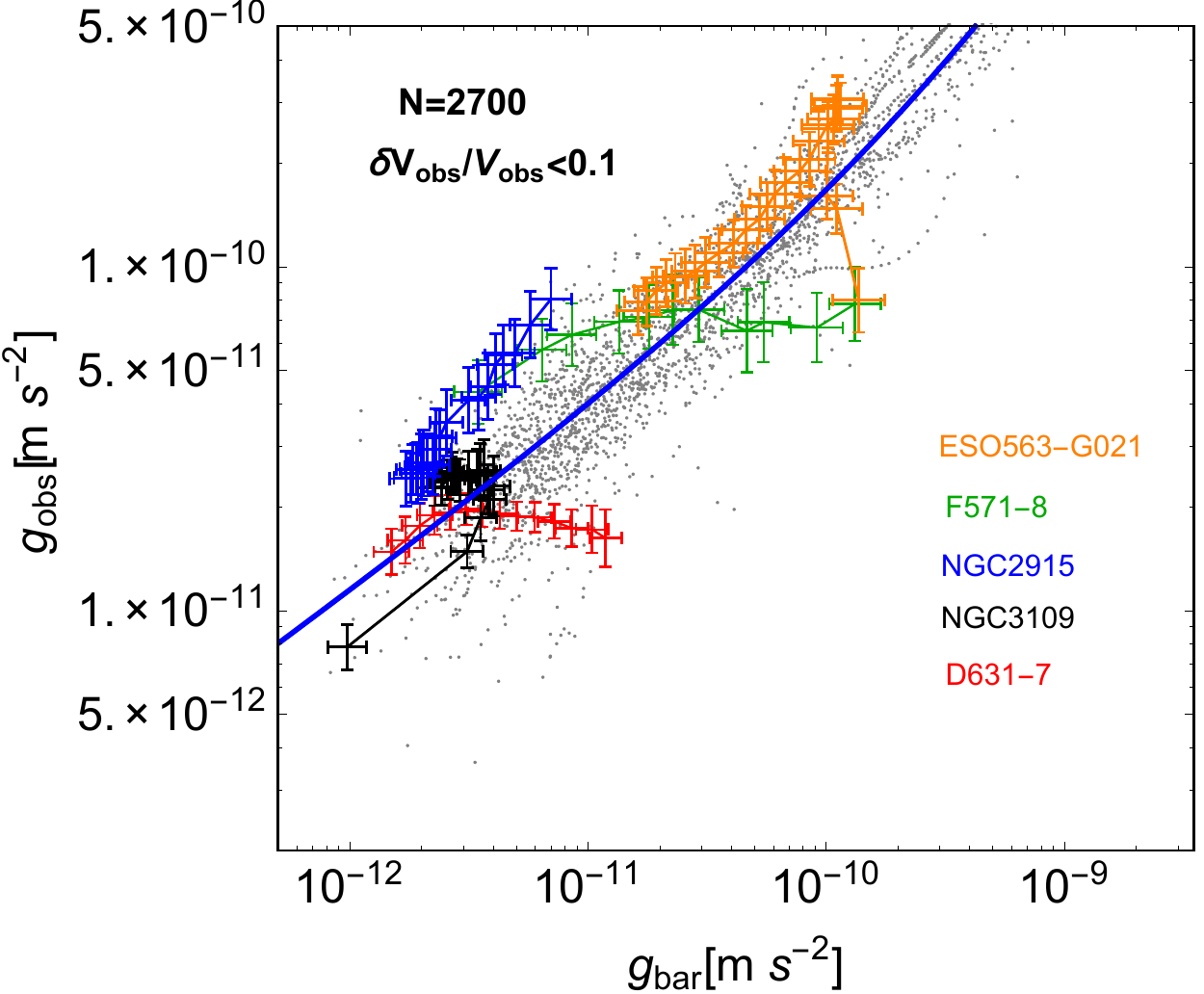}
	\includegraphics[width=0.35\textwidth]{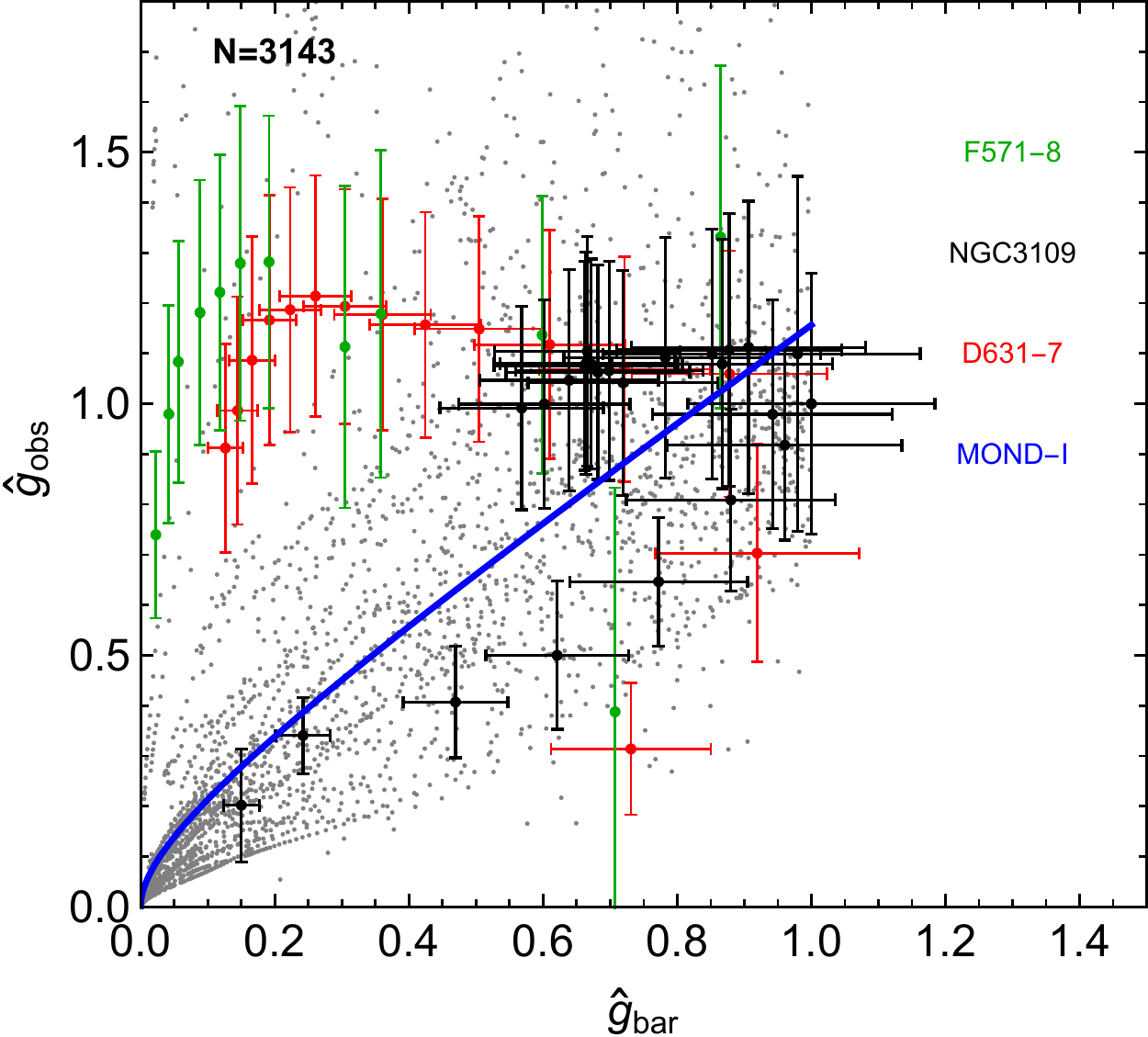}
		\includegraphics[width=0.35\textwidth]{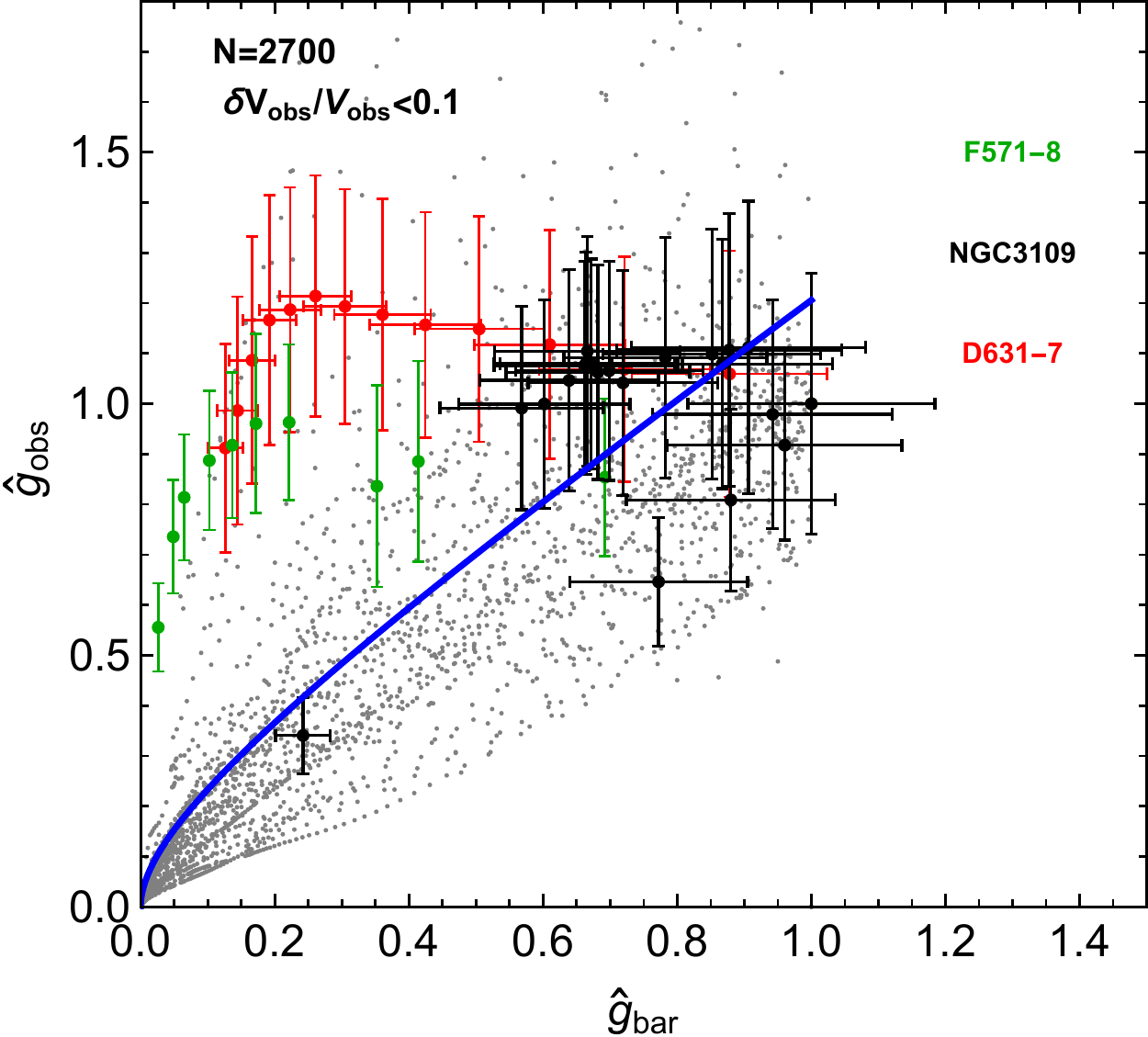}
	\caption
	{SPARC data in $g2$-space (upper panels) and $\hat{g}2$-space (lower panels). The full sets of data points without errors are shown as gray dots (3143 points on left panels) and also when imposing $\delta v_{\rm obs}/v_{\rm obs}<0.1$ (2700 points on right panels). On all panels we also show the prediction from MOND modified inertia with $g_0=1.2\times10^{-10} \frac{\rm{m}}{\rm{s}^2}$ (blue)  and individual galaxies with associated errors from  Eq.~(\ref{Eq:uncertaintiest}) are highlighted (color legend on figure). On the lower panels fewer individual galaxies are displayed for visual clarity. 
	}
	\label{Fig:Galaxies}
\end{figure}

We show (a part of) the collection of SPARC data in $g2$-space from these 152 galaxies in the top left panel of Fig.~\ref{Fig:Galaxies} (gray dots) across 3 orders of magnitude in $g_{\rm bar}$.  Also shown in the figure panel are the curves of individual galaxies with error bars that were highlighted in \cite{Petersen:2017klw}. These error bars include both random and systematic errors from Eq.~\ref{Eq:uncertaintiest}.  The blue line is the MOND modified inertia function in Eq.~\eqref{Eq:nufunction} with $g_0\simeq 1.2 \times 10^{-10} \frac{m}{s^2}$. This value of $g_0$ is the best fit value to the entire data set found in \cite{McGaugh:2016leg,Lelli:2017vgz} with the additional data requirement of $\delta v_{\rm obs}/v_{\rm obs}<0.1$. The top right panel shows the same figure with this requirement $\delta v_{\rm obs}/v_{\rm obs}<0.1$ imposed. Finally the bottom panels show the same data in the normalized $\hat{g}2$-space.

While the entire collection of data traces the MOND modified inertia curve, as observed and quantified in \cite{McGaugh:2016leg,Lelli:2017vgz}, it also appears that individual galaxies deviate significantly from this curve. 

\bigskip
In order to test the geometry of the data we therefore first consider 3 subsets of data points from the 152 galaxies, $N_{1,2,3}$. We denote the set of 152 data points with radii $ r_j=r_{\rm obs}$ by $N_{1}$ and the remaining 146 data points after first requiring $\delta {v_{\rm obs}}/v_{\rm obs}<0.1$ as in \cite{McGaugh:2016leg,Lelli:2017vgz} by $N_{2}$.
Computing the averages $\langle \hat{g}_{\rm obs, bar} \rangle $ on these sets we find that the $N_{1}$ and $N_{2}$ data sets yield 3$\sigma$ and more than 5$\sigma$ discrepancy respectively with the MOND modified inertia prediction $\langle \hat{g}_{\rm obs} (r_{\rm obs})\rangle_{\rm MI} $=1. The discrepancies with the prediction  $\langle \hat{g}_{\rm bar} (r_{\rm obs})\rangle_{\rm MI} $=1 are larger as summarized in table~\ref{Table:datasets1}.  
 \bigskip
   \bigskip
 \begin{table}[h!]
\begin{tabular}{ |p{2cm}||p{2cm}|p{5cm}|p{2cm}| p{3cm}| p{3cm}| }
 \hline
Data Sets & Galaxies &Data selection& Data points&$\langle\hat{g}_{\rm obs} \pm \delta \hat{g}_{\rm obs}  \rangle $ &$\langle\hat{g}_{\rm bar} \pm \delta \hat{g}_{\rm bar}  \rangle $   \\
 \hline
$N_{1}$ &   152 & $r_j=r_{\rm obs}$& 152 &$1.39\pm 0.12$&$0.83\pm 0.01$\\
$N_{2}$ &   152 & $r_j=r_{\rm obs}$ and $\delta v_{\rm obs}/v_{\rm obs}<0.1$& 146 &$1.12\pm 0.02$&$0.91\pm 0.01$\\
$N_{3}$ &   152 & $r_{{\rm obs, bar}+1}\geq r_j\geq  r_{{\rm obs,bar}-1}$& $\sim 400$ \footnote{402 points with $r_{{\rm obs}+1}\geq r_j\geq  r_{{\rm obs}-1}$ and 393 with $r_{{\rm  bar}+1}\geq r_j\geq  r_{{\rm bar}-1}$ where in some cases there is e.g. no  data point at radii below $r_{\rm bar}$. } &$1.23\pm 0.04$&$0.89\pm 0.01$\\
 \hline
\end{tabular}
\caption{Data sets $N_{{1,2}}$ on which the averages $\langle \hat{g}_{\rm obs,bar}(r_{\rm obs}) \rangle$ and errors are computed and $N_{3}$ on which  $\langle \hat{g}_{\rm obs,bar}(\Delta r_{\rm obs}) \rangle$ is computed. The MOND modified inertia predictions are 
$\langle \hat{g}_{\rm obs, bar} (r_{\rm obs})\rangle_{\rm MI} $=$\langle \hat{g}_{\rm obs, bar} (\Delta r_{\rm obs})\rangle_{\rm MI} =1$.
}
\label{Table:datasets1}
\end{table}
\bigskip
To improve the significance we consider a larger data set $N_{3}$ with a range  $r_j\in \Delta r_{\rm obs, bar}$ of points around $r_{\rm obs, bar}$ defined here via $r_{\rm obs, bar}+1\geq r_j\geq  r_{\rm obs,bar}-1$ . We compute  $\hat{g}_{\rm obs} (\Delta r _{\rm obs})$ as defined above using these points for each galaxy and finally the galaxy averages $\langle \hat{g}_{\rm obs,bar} (\Delta  r_{\rm obs}) \rangle $ over all galaxies with this data. Here we find more than 5$\sigma$ discrepancy from the MOND modified inertia prediction of unity with both the $\hat{g}_{\rm obs,bar}$ observables. The results are summarized in the last row in Table~\ref{Table:datasets1}.

The numbers summarized in Table~\ref{Table:datasets1} imply that MOND modified inertia does not correctly describe the SPARC data, even if the overall scatter around the fitting function~\eqref{Eq:nufunction} was found to be small in \cite{McGaugh:2016leg,Lelli:2017vgz}. 
To study the geometry of the SPARC data further we group the entire data set into points $\textcolor{magenta}{N_G^{+}}$ at $r\geq r_{\rm bar}$ and points $\textcolor{purple}{N_G^{-}}$  at $r<r_{\rm bar}$. We further divide the galaxies into 3 groups  $G_{1,2,3}$, motivated by the theoretical characterization in Table~\ref{Table:geometries}. 
Galaxies in $\textcolor{gray}{{{G_1}}}$ satisfy  $r_{\rm bar}=r_{\rm tot}$, galaxies in $G_2$ satisfy $r_{\rm bar}<r_{\rm tot}$,  and galaxies in $G_3$ satisfy $r_{\rm bar}>r_{\rm tot}$.
The set of data points in $G_1$ is $\textcolor{gray}{N_{G_{1}}}$  while we divide each set of data points within $G_{1,2}$ into subsets $\textcolor{orange}{N_{G_{2}}^+}$, $\textcolor{cyan}{N_{G_{3}}^+}$ with $r_j > r_{\rm bar}$ and $\textcolor{red}{N_{G_2}^-}$ and $\textcolor{blue}{N_{G_{3}}^-}$ with $r_j < r_{\rm bar}$. 
We summarize the datasets in table~\ref{Table:data} below. 
   \bigskip
 \begin{table}[h!]
\centering
\begin{tabular}{ |p{2cm}||p{3cm}|p{3cm}|p{2cm}| p{3cm}|  p{3cm}| }
 \hline
Data set & Galaxy selection&Data selection& Data points\\
 \hline
 \hline
 $\textcolor{magenta}{N_{G}^+}$ & \textcolor{magenta}{all} \textcolor{magenta}{(152)}&   \textcolor{magenta}{$r_j >r_{\rm bar}$ } &\textcolor{magenta}{2695 }\\
$\textcolor{purple}{N_{G}^-}$ & \textcolor{purple}{all} \textcolor{purple}{(152) }&   \textcolor{purple}{$r_j \leq r_{\rm bar}$} &\textcolor{purple}{ 296}  \\
$\textcolor{gray}{N_{G_{1}}}$  &$\textcolor{gray}{r_{\rm obs}=r_{\rm bar}}$ \textcolor{gray}{(29) }& \textcolor{gray}{all}& \textcolor{gray}{933}   \\
$\textcolor{orange}{N_{G_{2}}^+}$ &   \textcolor{orange}{$r_{\rm obs}>r_{\rm bar}$ } \textcolor{orange}{(86)}& \textcolor{orange}{$r_j > r_{\rm bar}$}&\textcolor{orange}{1179}  \\
$\textcolor{red}{N_{G_2}^-}$&   \textcolor{red}{$r_{\rm obs}>r_{\rm bar}$}  \textcolor{red}{(86) }& \textcolor{red}{$r_j \leq r_{\rm bar}$}& \textcolor{red}{140}  \\
$\textcolor{cyan}{N_{G_{3}}^+}$    & \textcolor{cyan}{$r_{\rm obs}<r_{\rm bar}$ }\textcolor{cyan}{ (37)}&\textcolor{cyan}{ $r_j > r_{\rm bar}$}&\textcolor{cyan}{764}\\
$\textcolor{blue}{N_{G_{3}}^-}$& \textcolor{blue}{$r_{\rm obs}<r_{\rm bar}$} \textcolor{blue}{(37)}& \textcolor{blue}{$r_j \leq r_{\rm bar}$ }&\textcolor{blue}{127}\\
 \hline 
\end{tabular}
\caption{Summary of galaxy group and corresponding data points used in Fig.~\ref{Fig:datascaled} to show the average geometry of data.}
\label{Table:data}
\end{table}
\bigskip
Within the above 7 galaxy data groups $N_{G}^{\pm}, N_{G_1}, N_{G_{2,3}}^{\pm}$ we bin the normalized baryonic accelerations $\hat{g}_{\rm bar}(r_j)$ in 4 bins of width $\hat{g}_{\rm bar, k }-\hat{g}_{\rm bar, k-1 }=\Delta \hat{g}_{\rm bar}=0.25$ with  $k=1,...,4$ and compute the average values $\langle \hat{g}_{\rm bar, obs} \rangle_{N_{G_{i,k}}^{\pm}}$ and associated errors $\delta \langle \hat{g}_{\rm bar, obs} \rangle_{N_{G_{i,k}}^{\pm}}$ discussed in the appendix.

\begin{figure}[htp!]
	\centering
	\includegraphics[width=0.4\textwidth]{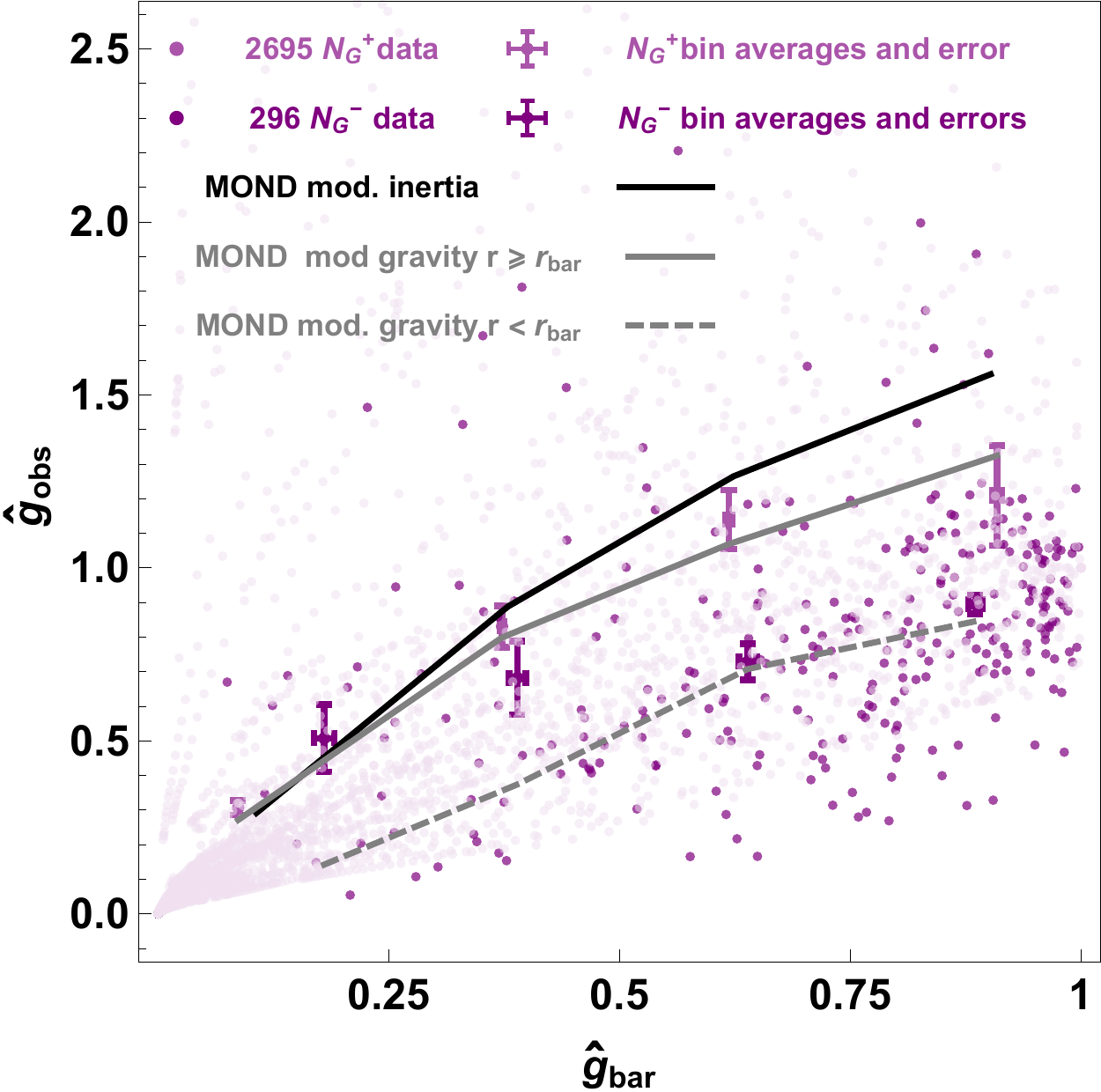}
	\includegraphics[width=0.4\textwidth]{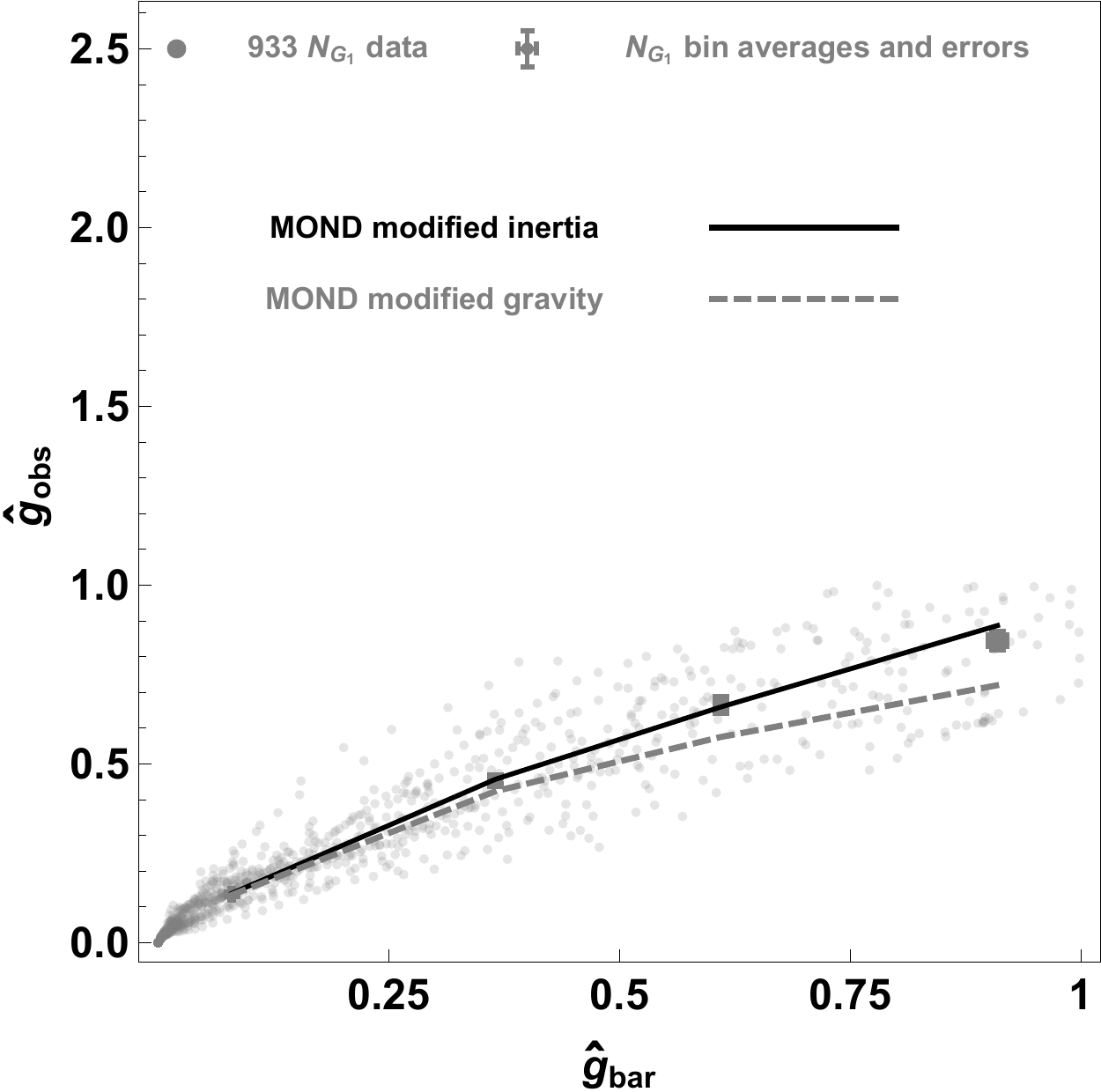}
	\includegraphics[width=0.4\textwidth]{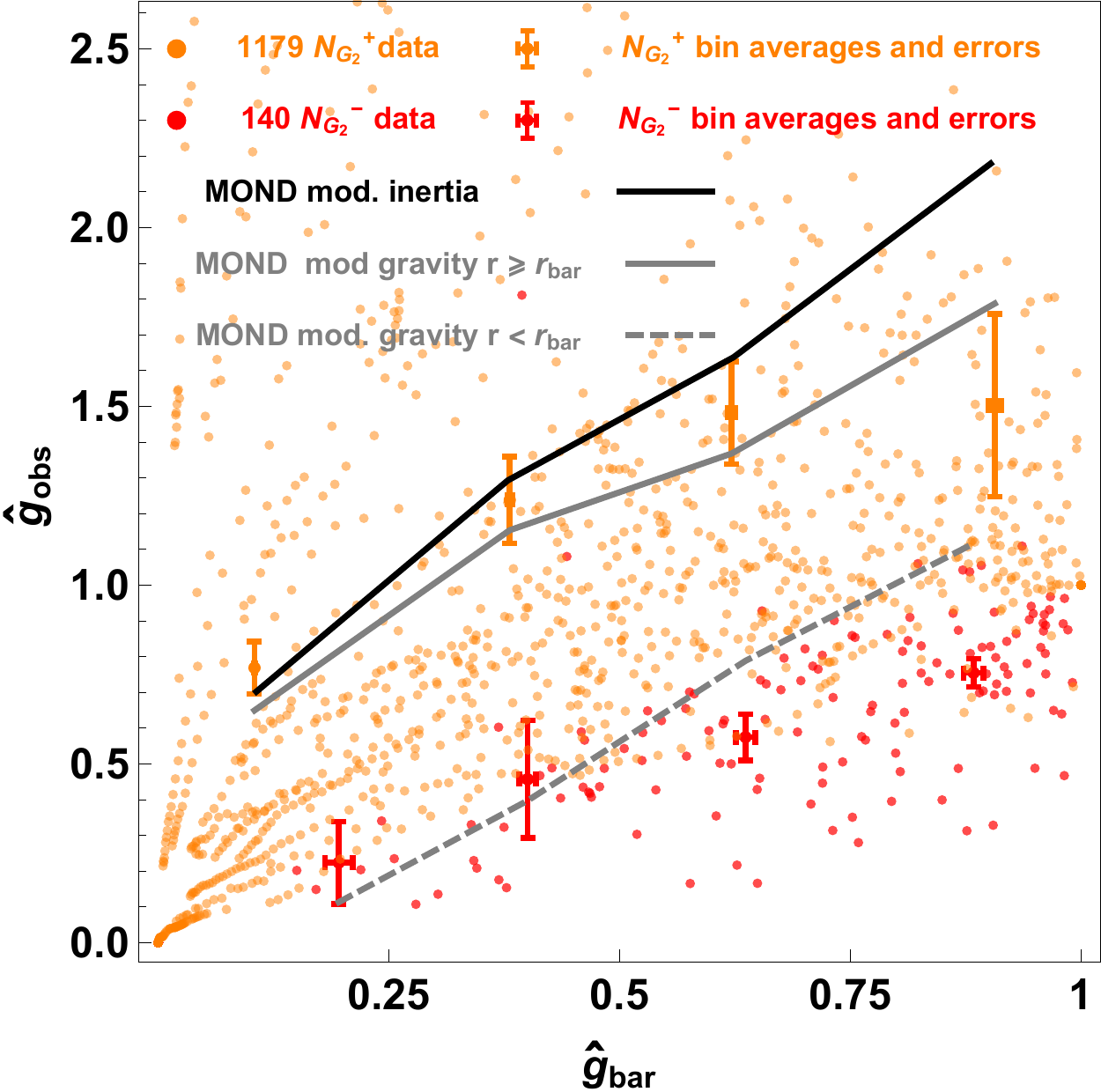}
	\includegraphics[width=0.4\textwidth]{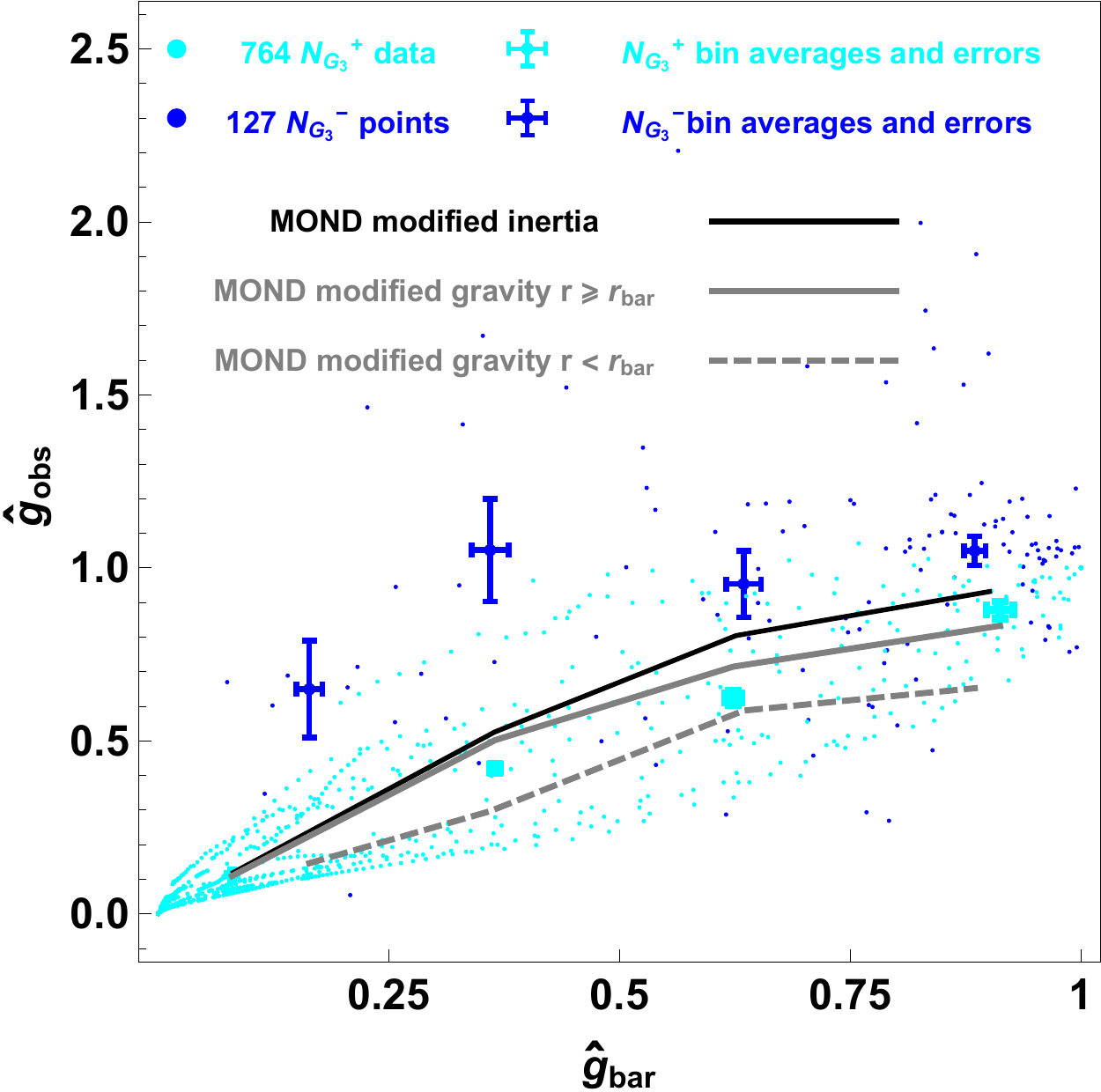}
	\caption
	{SPARC rotation curve data and data averages with errors in the normalized $\hat{g}2$-space. 
	\newline
	{\it Top left panel}: The full SPARC data set (shown without errors) divided into points in $N_{G}^+$ with $r>r_{\rm bar}$ (light purple) and those in $N_{G}^-$ with $r<r_{\rm bar}$ (purple). Also shown are the average data values and their errors computed within the 4 $\hat{g}_{\rm bar}$ bins in $N_{G}^{\pm}$ (light purple and purple error bars) as discussed in the text. Finally we show the averaged prediction from MOND MI (black curve) and MOND MG for $r>r_{\rm bar}$ (gray solid) and for $r>r_{\rm bar}$ (gray dashed ).\newline
	{\it Top right panel}: The same as top left but for all data in $N_{G_1}$ (galaxies where $r_{\rm obs}=r_{\rm bar}$) without distinguishing between $r>r_{\rm bar}$  or  $r<r_{\rm bar}$. \newline
	{\it  Bottom left panel}: The same as top left but for data in $N_{G_2}$ (galaxies where $r_{\rm obs}>r_{\rm bar}$). \newline
	{\it Bottom right panel}: The same as top left but for data in $N_{G_3}$ (galaxies where $r_{\rm obs}<r_{\rm bar}$).
			}
	\label{Fig:datascaled}
\end{figure}

We show the data groups $N_G^{\pm}, N_{G_1}, N_{G_{2,3}}^{\pm}$ together with the binned averages of each corresponding data set in Fig.~\ref{Fig:datascaled}. On all 4 panels the solid black line is the MOND modified inertia prediction while the solid and dashed gray lines are the predictions from the Bekenstein-Milgrom MOND modified gravity approximation at radii above and below $r_{\rm bar}$. We keep the discussion below qualitative as we have already presented the quantitative discrepancy with MOND modified inertia and because our treatment of MOND modified gravity relies on the approximation for purely disk galaxies in \cite{Brada:1994pk}.

The top left panel shows data from the full group of SPARC galaxies, equivalent to Fig~\ref{Fig:Galaxies}, but with data divided into the two groups $N_G^{\pm}$. The data (light purple and purple dots) is seen to display the geometry characterized by $r_{\rm tot}>r_{\rm bar}$ in table~\ref{Table:geometries} on average. MOND modified inertia (black line) is a good description of the average values of $\textcolor{magenta}{N_G^{+}}$ (light purple points with errors) but not in $\textcolor{purple}{N_G^{-}}$ (purple points with errors)  at large accelerations. Also the panel shows a large overall spread in data in $\hat{g}2$-space compared to the data errors on the averages. 
MOND modified gravity (solid gray for $r\geq r_{\rm bar}$ and dashed gray line for $r<r_{\rm bar}$) is a better description of data except for points at $r<r_{\rm bar}$ with small accelerations. The panel also shows a large overall spread in data in $\hat{g}2$-space compared to the data errors on the averages.

The top right panel displays the same quantities but for the data set $\textcolor{gray}{N_{G_1}}$ where galaxies have  $r_{\rm tot}=r_{\rm bar}$.  Here MOND modified inertia is a very good description of the averaged data --- which by selection only samples radii $r\geq r_{\rm bar}$. Both the average measurement error and the spread in data is smaller than for the full data set (both at $r<r_{\rm bar}$ and at $r\geq r_{\rm bar}$) on the left panel.

The bottom left panel is for the $\textcolor{orange}{N_{G_2}^{+}}$ and $\textcolor{red}{N_{G_2}^{-}}$ data sets with $r_{\rm tot}>r_{\rm bar}$ which is true for most of the galaxies (86) and these galaxies are driving the overall geometry and and data spread seen in the top left panel. Despite this spread and the greater average errors there is a clear difference between the two data sets $\textcolor{orange}{N_{G_2}^{+}}$ and $\textcolor{red}{N_{G_2}^{-}}$ with MOND modified inertia a poor description of 
$\textcolor{red}{N_{G_2}^{-}}$ data. 
 Again the MOND modified gravity prediction is clearly a better match to the data, but as opposed to the full data set in the top left panel, it is now the data at $r<r_{\rm bar}$ and large accelerations that yields the biggest deviations.

Finally the right hand panel shows the results for the $\textcolor{cyan}{N_{G_3}^{+}}$ and $\textcolor{blue}{N_{G_3}^{-}}$ data sets with $r_{\rm tot}<r_{\rm bar}$.  
Here only the spread and errors of the $\textcolor{blue}{N_{G_3}^{-}}$ set is big as compared to the $N_2$ set, with MOND modified inertia model match to the average values of both data sets.  Inevitably the MOND modified gravity approximation is also a poor match to the average values of the  $\textcolor{blue}{N_{G_3}^{-}}$ as the MOND modified gravity approximation always leads to $r_{\rm tot}>r_{\rm bar}$

Again we do not here quantify the deviations of the MOND modfied gravity approximation, as this approximation was developed for an infinitely thin disk galaxy geometry  \cite{Brada:1994pk} and also does not take into account the external field effect \cite{1983ApJ...270..365M} in MOND modified gravity, which might be important for some non-isolated galaxies, see e.g. the recent discussion of the 'dark matter less' dwarf galaxy NGC-1052-DF2 and MOND \cite{vanDokkum:2018vup,Famaey:2018yif}. 
The analysis does show that the disagreement is driven by the majority of galaxies exhibiting geometries with $r_{\rm tot}>r_{\rm bar}$ but it is offset by a minority of galaxies exhibiting $r_{\rm tot}<r_{\rm bar}$. This, together with the fact that most data is measured at $r > r_{\rm bar}$, means that as a whole the SPARC rotation curve data exhibits moderate and Gaussian residuals around the function~\eqref{Eq:nufunction} as found in \cite{McGaugh:2016leg}. This however does not reflect the average geometry of the rotation curves. 
Our analysis therefore highlights the need to further study MOND Modified gravity models, beyond the MOND modified inertia models most often used in the literature, in order to establish if MOND can account for rotation curve data.

\newpage
\section{Summary}
\label{Sec:Summary}
\noindent We have shown the $g2$-space geometry of selected MOND and DM models for disk galaxies with exponential mass densities for the visible baryonic mass distribution  in Fig.~\ref{Fig:MONDfunctions} --- these are MOND modified inertia and an approximate description of Bekenstein-Milgrom MOND modified gravity models as well as DM models with NFW and quasi-isothermal DM density profiles. 

We have classified the $g2$-space geometry of these models in Figs.~\ref{Fig:MONDfunctions} and ~\ref{Fig:MONDfunctionsscaled} using global characteristics: The location of the maximum acceleration due to the baryonic matter and the maximum of the total predicted acceleration, $r_{\rm bar}$ and $r_{\rm tot}$, whether the curve is closed or open and the area of the closed cuves $\mathcal{A}(\mathcal{C})$. MOND modified inertia models, DM models with NFW profiles and DM models with quasi-isothermal profiles can be organized in distinct categories according to these global characteristics, while MOND modified gravity models in the approximation used is degenerate with DM models with quasi-isothermal profiles as summarized in table~\ref{Table:geometries}.

Rotation curve data may also be organized according to this classification. Applying this classification to rotation curve data from the SPARC data base we find that MOND modified inertia, independent of the specific interpolation function used, is in disagreement with the data at more than 5$\sigma$. A previous analysis finding disagreement between MOND modfied inertia and SPARC data was presented in \cite{Petersen:2017klw}. 
In the current analysis we have considered ratios of accelerations $\hat{g}_{\rm bar, obs}(r)\equiv g_{\rm bar, obs}(r)/g_{\rm bar, obs}(\rm r_{\rm bar})$ with respect to some reference acceleration, here chosen as  $g(\rm r_{\rm bar})$ in order to reduce the systematic uncertainties in data stemming from galaxy inclination angles $i$ and distances $D$ on ${g}_{\rm obs}$ as well as mass to light ratios $\Upsilon_{\rm disk, bulge}$ on ${g}_{\rm bar}$. 
If there is a strong radial dependence of these quantities within individual galaxies, and/or between galaxies this can still affect our results. However, changing the conclusion that MOND modified inertia models do not fit the data would require significant radius variations from $r_{\rm bar}$ to $r_{\rm tot}$. A detailed study of this is beyond the scope of this paper, but e.g. a monotonically decreasing dependence of mass to light ratios with radius \cite{Portinari:2009ap,deDenus-Baillargeon:2013uga} will not change our result that MOND modified inertia is not in agreement with data.
  
We have presented the rotation curve data from the SPARC data base organized according to the relative location of $r_{\rm bar, tot}$ in $\hat{g}2$-space in table~\ref{Table:data} and Fig.~\ref{Fig:datascaled}. In addition to the quantitative results on MOND modified inertia, these figures establish qualitatively that subsets of galaxies display different geometric characteristics and neither MOND modified inertia nor MOND modified gravity describe all data subsets.  
If all data is joined together a fit to MOND modified inertia with gaussian errors and moderate scatter can be obtained \cite{McGaugh:2016leg} since the average data from the data sets $\textcolor{red}{N_{G_2}^{-}}$   and $\textcolor{blue}{N_{G_3}^{-}}$ 
deviate in opposite directions from the MOND modified inertia prediction and since most data points are measured at $r>r_{\rm bar}$ where deviations from MOND modified inertia are not as significant. Since the global geometrical characteristics of the other considered models, both MOND modified gravity (in the approximation employed), DM with isothermal density profile and DM with NFW density profile, differ from MOND modified inertia exactly for data points at $r<r_{\rm bar}$ it is important to investigate these separately. 

In summary we find that MOND modified inertia models, frequently used to fit rotation curve data, are not in agreement with data, while further study of MOND modified gravity models would be required to establish those as a viable explanation of data.  
Further we find that the detailed geometry in $g2$-space is useful to probe different DM density distributions, with e.g. only a minority of galaxies exhibiting the global characteristics of NFW profiles. This latter conclusion is  well known in the guise of the cusp-core problem. However the $g2$-space analysis makes it apparent how in particular future improvements in rotation curve data at small radii is extremely useful in probing the DM density profile.  This may yield new insights on the required particle physics characteristics of DM, e.g. DM self interactions. More generally the $g2$-space analysis offers a very useful and striking characterization of models for the missing mass problem.

\bigskip
{\bf Acknowledgments:}
We thank W-C. Huang and I. Shoemaker for comments on the draft. 
The authors acknowledge partial funding from The Council For Independent Research, grant number DFF 6108-00623. The CP3-Origins center is partially funded by the Danish National Research Foundation, grant number DNRF90.

\appendix

\section{Error treatment}
\label{sec:app}
\noindent In this appendix we review calculation of errors used throughout the paper.
The errors $\delta g_{\rm obs,bar}(r_j)$ on individual $g_{\rm obs, bar}(r_j)$ points are given in the main text in Eq.~\eqref{Eq:uncertaintiest}.  The scaling of radius $r$, baryonic velocities $v_k$, with $k={\rm disk, bulge, gas}$, and observed velocities $v_{\rm obs} $  under a change of galaxy distance $D$ and inclination angle $i$ are 
\begin{align}
r &\to r' = \frac{D'}{D} r , \quad v_k \to v_k'=\sqrt{ \frac{D'}{D}} v_k , \quad v_{\rm obs} \to v'_{\rm obs} =\frac{\sin(i')}{\sin(i)} v_{\rm obs}  ; \quad
\end{align}
Therefore $\delta g_{\rm bar}$ is independent of distance $D$ and inclination angle $i$ as discussed in e.g. \cite{Li:2018tdo}, with the resulting scalings of $g_{\rm obs, bar}(r_j)$ being 
\begin{align}
 g_{\rm bar} &\to  g_{\rm bar}'= g_{\rm bar}  , \quad 
g_{\rm obs} \to g_{\rm obs}'=\frac{D}{D'}\frac{\sin(i')^2 }{\sin(i)^2} g_{\rm obs}  
\end{align}

Once we form the ratios $\hat{g}_{\rm bar, obs}$ then also $\hat{g}_{\rm obs}$ is independent of distance $D$ and inclination angle $i$ such that under a change of distance $D$ and angle $i$ we have 
\begin{align}
 \hat{g}_{\rm bar}'= \hat{g}_{\rm bar}  , \quad \hat{g}_{\rm obs} \to \hat{g}_{\rm obs}'=\hat{g}_{\rm obs}
\end{align}
We include the systematic uncertainty in $\hat{g}_{\rm bar}$ from the mass to light ratios $\Upsilon_{\rm disk, bulge}$ via propagation of errors including covariance, such that  
\begin{align}
{\rm Cov}(f_k,f_l)& = \sum_a \sum_b  \frac{\partial f_k }{\partial x_a} \frac{\partial f_l }{\partial x_b}  {\rm Cov}(x_a,x_b) \nonumber
\\
 x_{a}&= \{ \Upsilon_{\rm disk}, \Upsilon_{\rm bulge} , v_{\rm gas}(r_j)\} \nonumber
\\   
f_{k}&= \{ g_{\rm bar}(r_j) , g_{\rm obs}(r_j) \}  
\label{Eq:coverror2}
\end{align}
where ${\rm Cov(x_a,x_a)}=\delta x_a^2$ is the error of $x_a$, ${\rm Cov(x_a,x_b)}=0$ for uncorrelated errors $x_{a,b}$, ${\rm Cov(x_a,x_b)}=\delta x_a \delta x_b$ for fully correlated errors $x_{a,b}$ and similar for the functions $f_{k,l}$.  The functions  $f_{k,l}$ are the entire set of accelerations $g_{\rm bar, obs}(r_j)$ and from this covariance matrix we find the errors on  $\hat{g}_{\rm bar, obs}(r_j)$ and errors on averages   $\langle \hat{g}_{\rm bar, obs}(r_j) \rangle$ which we discuss explicitly below. 
First the errors $\delta \hat{g}_{\rm bar, obs} (r_j)$ following from Eq.~\eqref{Eq:coverror2} are 
\begin{align}
\delta \hat{g}_{\rm obs} (r_j) &=  \hat{g}_{\rm obs}  (r_j) \sqrt{ \left(\frac{2 \delta v_{\rm obs} (r_j)}{ v_{\rm obs} (r_j)}\right)^2 +  \left(\frac{2 \delta v_{ \rm obs }(r_{\rm bar }) }{ v_{ \rm obs }(r_{\rm bar })}\right)^2 }  \quad {\rm for} \quad r_j \neq r_{\rm bar}  \nonumber
 \\ 
\delta \hat{g}_{\rm bar}  (r_j)&=  \hat{g}_{\rm bar}  (r_j)
\sqrt{ \left(\frac{2  v_{\rm gas}(r_j)  \delta v_{\rm gas}(r_j)}{ v_{\rm bar}(r_j)^2}\right)^2+  \left(\frac{2 v_{ \rm gas }(r_{\rm bar })  \delta v_{ \rm gas }(r_{\rm bar }) }{ v_{ \rm bar }(r_{\rm bar })^2 }\right)^2  + \left( \Delta \Upsilon (r_j) \right)^2} , \quad {\rm for} \quad r_j \neq r_{\rm bar} , \nonumber
\\
 \Delta \Upsilon (r_j)& = \sum_{k={\rm disk, bulge}} \delta \Upsilon_k  \left( \frac{v_k^2(r_j)}{v_{\rm bar}^2(r_j)} - \frac{ v_{k}^2(r_{\rm bar })}{v_{\rm bar }^2(r_{\rm bar })} \right) 
\label{Eq:haterrors}
\end{align}
while $\delta \hat{g}_{\rm obs} (r_{\rm bar}) =\delta \hat{g}_{\rm bar} (r_{\rm bar}) =0$.
It follows from Eq.~\eqref{Eq:haterrors} that $\hat{g}_{\rm bar}(r_j)$ is insensitive to the systematic uncertainties in $\delta \Upsilon_k$ near $r_{\rm bar}$ by construction,
 where we are particularly interested in the geometry.

In summary the ratios $\hat{g}_{\rm bar,obs} $ eliminate the systematic uncertainties in galaxy distance and disk inclination and significantly reduce that from mass to light ratios. These three sources of systematic uncertainties were found to be the dominant sources of scatter in previous analysis of SPARC data \cite{Li:2018tdo}. 
We have checked explicitly that the error $\Delta \Upsilon$ on $\hat{g}_{\rm bar}(r_j)$ is indeed small and while we keep it in all error calculations this means we can take $\hat{g}_{\rm obs,bar}(r_j)$ values from different galaxies to be uncorrelated even if $\delta \Upsilon_k$ are correlated between different galaxies --- of course if mass to light ratios between different galaxies vary randomly then so do $\hat{g}_{\rm obs,bar}(r_j)$ regardless of this residual error being small. 
5while values within a galaxy are still correlated via the same normalization points $\hat{g}_{\rm obs,bar}(r_{\rm bar})$. 

\subsection{Errors on averages}
From the above errors on individual $\hat{g}_{\rm obs,bar}(r_j)$ the points $\hat{g}_{\rm bar, obs} (r_{\rm obs})$ over all galaxies are uncorrelated and their averages and errors presented in Table~\ref{Table:datasets1} are simply given from Eq.~\ref{Eq:haterrors} by
\begin{align}
\langle \hat{g}_{\rm bar, obs} (r_{\rm obs})\rangle_{N_{1,2}}&=
\frac{1}{ N_{1,2}} \sum_{ j\in N_{1,2}}  \hat{g}_{\rm bar, obs}(r_j),  \quad 
\delta \langle  \hat{g}_{\rm bar,obs} \rangle_{N_{1,2}}=\frac{1}{ N_{1,2}} \sqrt{  \sum_{r_j \in N_{1,2}} \delta \hat{g}_{\rm bar,obs}^2 } .
\end{align}

The averages and errors on the average of $\hat{g}_{\rm obs}$ (and similarly with $\hat{g}_{\rm bar}$) for points within a galaxy $G$ can be written as 
\begin{align}
\langle \hat{g}_{\rm obs} \rangle_G&=
\frac{1}{ N_G } \sum_{ j\in G}  \hat{g}_{\rm obs}(r_j),  \quad  \langle \delta \hat{g}_{\rm obs} \rangle_{G}=\frac{1}{ N_{G} }\sqrt{  \sum_{r_j \neq r_{\rm bar}}   \left( \hat{g}  (r_j) \frac{2 \delta V_{\rm obs} (r_j)}{ V_{\rm obs} (r_j)} \right)^2 + \left( \sum_{r_j \neq r_{\rm bar}}    \hat{g}  (r_j)  \frac{2 \delta V_{ \rm obs }(r_{\rm bar }) }{ V_{ \rm obs }(r_{\rm bar })}\right)^2}
\label{Eq:errorgal}
\end{align}
where the error will typically be dominated by the last term, which is $O(1)$ in the number of points $N_G$ while the first term is $O(1/\sqrt{N_G})$  due to the single normalization point in the denominator. To improve on this we also employ the average ${g}_{\rm bar,obs}(\Delta r_{\rm bar})$  for the last results with data set $N_3$ in table~\ref{Table:datasets1} such that
\begin{align}
\langle \hat{g}_{\rm bar, obs} \rangle_{N_{3}}&= \frac{1}{ N_G} \sum_{N_{G}} 
\frac{N_{\Delta_{\rm bar,G}}}{N_{\Delta_{\rm obs,G}}} \  \hat{g}_{\rm bar, obs}( \Delta r_{\rm obs})
,  \\ 
\delta \langle  \hat{g}_{\rm bar,obs} \rangle_{N_{3}} &=\frac{1}{N_G}  \sqrt{ \sum_{N_{G}} \left( \frac{N_{\Delta_{\rm bar}}}{N_{\Delta_{\rm obs}} }  \delta \hat{g}_{\rm bar, obs}( \Delta r_{\rm obs}) \right)^2 } 
\end{align}
where again $N_G$ is the number of galaxies used in the average  $\Delta r_{\rm bar,obs}$ are the intervals around $r_{\rm bar,obs}$ and $N_{\Delta_{\rm obs,G}}$ are included to correct for cases when either $\Delta r_{\rm bar}$ or  $\Delta r_{\rm obs}$ contain less than 3 points.

Finally the errors on the binned averages over the points $N_{i,k}^{\pm}$ in Fig.~\ref{Fig:datascaled} are computed by first computing the error on the points $G\cap N_{i,k}^{\pm}$ in $N_{i,k}^{\pm}$ from a given galaxy $G$ as in Eq.~\eqref{Eq:errorgal} which then are uncorrelated between galaxies such that the weighted errors are:
\begin{align}
\langle \hat{g}_{\rm bar, obs} \rangle_{N_{i,k}^{\pm}}&=\frac{1}{ N_{i,k}^{\pm} } \sum_{j\in N_{i,k}^{\pm} }   \hat{g}_{\rm bar, obs} (r_j)\\
\delta \langle \hat{g}_{\rm bar, obs} \rangle_{N_{i,k}^{\pm}} &=\frac{1}{ N_{i,k}^{\pm} } \sqrt{\sum_{ G} N_{G\cap N_{i,k}^{\pm}} \delta \langle  \hat{g}_{\rm obs} \rangle_{G\cap N_{i,k}^{\pm}  }^2  }
\end{align}

\newpage

\bibliography{rot}
\bibliographystyle{hunsrt}

\end{document}